\documentclass{aa}  
\usepackage{graphicx}
\usepackage{longtable}
\usepackage{multirow,multicol}
\usepackage{float}
\usepackage{tabularx}
\usepackage{cuted}
\usepackage{textcomp}
\usepackage{amsmath}
\usepackage{gensymb}
\usepackage{comment}

\usepackage{natbib,twoopt}
\usepackage[breaklinks=true]{hyperref} 
\bibpunct{(}{)}{;}{a}{}{,}             
\makeatletter
  \newcommandtwoopt{\citeads}[3][][]{\href{http://adsabs.harvard.edu/abs/#3}%
    {\def\hyper@linkstart##1##2{}%
     \let\hyper@linkend\@empty\citealp[#1][#2]{#3}}}
  \newcommandtwoopt{\citepads}[3][][]{\href{http://adsabs.harvard.edu/abs/#3}%
    {\def\hyper@linkstart##1##2{}%
     \let\hyper@linkend\@empty\citep[#1][#2]{#3}}}
  \newcommandtwoopt{\citetads}[3][][]{\href{http://adsabs.harvard.edu/abs/#3}%
    {\def\hyper@linkstart##1##2{}%
     \let\hyper@linkend\@empty\citet[#1][#2]{#3}}}
  \newcommandtwoopt{\citeyearads}[3][][]%
    {\href{http://adsabs.harvard.edu/abs/#3}
    {\def\hyper@linkstart##1##2{}%
     \let\hyper@linkend\@empty\citeyear[#1][#2]{#3}}}
\makeatother

\usepackage{txfonts}
\begin{document}

   \title{Recovering cosmological parameters from the mock\\gravitational wave data of the Einstein Telescope}


   \author{Pinaki Roy
          \inst{1}\fnmsep\thanks{\email{p.roy@uw.edu.pl}}
          \and
          Tomasz Bulik\inst{1}
          }

   \institute{Astronomical Observatory, University of Warsaw, Al. Ujazdowskie 4, 00-478 Warsaw, Poland.}
         
   \date{Received MMMMM DD, YYYY; accepted MMMMM DD, YYYY}

 
  \abstract
   {\emph {Context.} Einstein Telescope (ET) is a third-generation gravitational wave (GW) detector with tenfold better sensitivity compared to the advanced LIGO detectors. It will be capable of observing copious stellar mass binary black hole mergers upto a redshift of 10 which will make it especially useful for cosmography.\\
   \emph {Aim.} We generate a mock gravitational wave event catalog for the Einstein Telescope and show the recoverability of either the Hubble constant ($H_0$) or the matter density parameter ($\Omega_{\rm m}$).\\
   \emph{Methods.} We present a simple, effective and fast technique for inferring $H_0$ (or $\Omega_{\rm m}$) using the intrinsic chirp mass spectrum of black hole binaries, and investigate the efficacy of the method assuming the standard model of cosmology.\\
   \emph{Results.} If only $H_0$ has to be constrained, we find that at least one year of ET's observation will be required to achieve 1\% uncertainty. With the same amount of observation, $\Omega_{\rm m}$ can be constrained to within 4\% uncertainty.\\
   \emph{Conclusion.} With ET operating as a standalone instrument, we show that the GW spectral sirens detected by it can constrain the Hubble constant.}

   \keywords{Gravitational waves; Stars: neutron, black holes; Methods: data analysis}

   \maketitle
%

\section{Introduction}

   The advent of gravitational wave (GW) astronomy with second-generation detectors like LIGO and Virgo has unveiled a population of merging compact binaries and provided groundbreaking multi-messenger insights. The future Einstein Telescope (ET), a third-generation GW telescope \citep{2010CQGra..27s4002P}, promises a quantum leap in sensitivity. 
   The annual detection rates based on the sensitivity for ET's triangular design \citep{2011CQGra..28i4013H}, namely ET-D, are expected to be $\sim 10^5-10^6$ binary black hole mergers (roughly 1 event every 100 seconds) and $\sim 7\times 10^4$ binary neutron star mergers (roughly 2 events every 15 minutes) \citep[see][]{2012PhRvD..86l2001R,2014PhRvD..89h4046R,2019JCAP...08..015B}. These high detection rates will make ET a powerful tool for population studies as well as cosmography.
   
   
   A key limitation for a single observatory such as the current GW detectors is the difficulty in localizing sources and breaking degeneracies between the intrinsic chirp mass of the GW source (merging binary) and the source redshift. \cite{2021PhRvD.104d3014S} showed how to address this challenge for the ET-D design by utilizing the antenna response of ET's three detectors. By analyzing the ratios of signal-to-noise and phase differences between the detectors, one can constrain the sky location, inclination, and polarization of the source. This, in turn, allows us to infer the luminosity distance of the source. From this, one can estimate the source redshift by assuming a cosmological model. However, if the cosmology to be used is uncertain, one can avail the vast catalog of GW events detected by ET to eliminate this uncertainty.

   To break the degeneracy between the source redshift and the intrinsic chirp mass of the binary, one can obviously exploit any associated electromagnetic (EM) transient to constrain the sky position with high accuracy \cite{1986Natur.323..310S,2017PhRvL.119p1101A}, subsequently identify the host galaxy and get the redshift distance. Such events are called bright sirens. GW events without an EM counterpart are referred as dark sirens. In such scenarios, one has to resort to statistical techniques to lift the degeneracy. The most popular of these is the galaxy survey method in which one cross-correlates the sky localization volume with galaxy catalogs in order to identify potential host galaxies \citep[see, e.g.][]{2012PhRvD..86d3011D,2018Natur.562..545C,2019ApJ...871L..13F,2023AJ....166...22G} and thus, get the probable redshift values of the source.

   A novel way is the spectral siren method, which exploits features in the source-frame mass distribution (also called the mass spectrum) of the merging binary systems, and use a large catalog of events to statistically break the mass-redshift degeneracy \citep[see, e.g.][]{2012PhRvD..85b3535T,2019ApJ...883L..42F,2021ApJ...908..215Y,2021PhRvD.104f2009M}. This approach facilitates an independent inference of cosmological parameters. The features in the mass spectrum arise from various parameters governing the binary evolution process \citep[see][]{2023pbse.book.....T}.

   We aim to leverage the intrinsic chirp mass distribution of binary black hole systems and ET's detection and localization capability to infer the Hubble constant and the matter density parameter. Our work differs from similar recent studies \citep[see, e.g.][]{2022PhRvL.129f1102E,2025ApJ...985L..33R,2025PhRvD.111l3535C,2026arXiv260217756T}. We include low-SNR and high-redshift events in the mock data. We do not distinguish between NS-BH and BH-BH binaries in our analysis. To the best of our knowledge, the KL divergence method we adopt has not been applied to this cosmography with spectral siren problem so far. Our approach can also be applied to the full mass spectrum but we reserve that exercise for a future study.
   
   
   This paper is organized as follows. In Section 2, we lay out the steps to build the mock GW data for ET. In Section 3, we present our results obtained with the mock data. In Section 4, we list the takeaways from our study and future plans.
   

\section{Method}

   
%

   We first describe the scheme for creating the mock data catalog for ET followed by a description of how we use the event catalog for cosmography. We have assumed standard model of cosmology (SMOC) throughout the analysis. In the minimal 6-parameter $\Lambda$CDM model, the curvature density parameter, $\Omega_k=0$, the radiation density parameter, $\Omega_{\rm rad}=0$ and the dark energy equation of state parameter, $w=-1$. Then, the matter density parameter ($\Omega_{\rm m}$) and the dark energy density parameter ($\Omega_\Lambda$) sum to unity, i.e. $\Omega_\text{m}+\Omega_\Lambda=1$.

   In this work, we first set ourselves the goal of recovering $H_0$ for two cases with different injected $H_0$ values. The two values lie at the typical extremes of the measured range of values. For Case I, we have used 67.3 km s$^{-1}$ Mpc$^{-1}$, whereas for Case II, we have used 73.5 km s$^{-1}$ Mpc$^{-1}$. In both these cases, we have set $\Omega_{\rm m}=0.3$. Then, we aim to recover $\Omega_{\rm m}$ keeping $H_0$ fixed at 70 km s$^{-1}$ Mpc$^{-1}$. For this, we prepare two more cases viz. Case III: $\Omega_{\rm m}=0.27$ and Case IV: $\Omega_{\rm m}=0.32$. Further, we consider additional cases with the injected values: $H_0=70$ km s$^{-1}$ Mpc$^{-1}$ and $\Omega_{\rm m}=0.3$, assuming some systematic uncertainty in the fixed parameter.

   We generate compact binary population using the publicly available binary population synthesis code \texttt{COMPAS} \citep{2022ApJS..258...34R} \texttt{v03.07.02} which can rapidly evolve isolated binaries through all the important stages of evolution such as the Roche lobe overflow (RLOF), mass transfer, supernova and common envelope. We evolve 1 million zero-age main sequence (ZAMS) binaries with Kroupa mass function and thermal eccentricity distribution for each of the 76 different metallicity values:
   $Z=0.0001-0.0020$ (in steps of 0.0001), $Z=0.0025-0.0300$ (in steps of 0.0005).

   As in \texttt{COMPAS}, so do we, in our analysis, identify a compact object as a neutron star if it is below 2.5 $M_\odot$ and as a black hole if it is above this mass limit. We choose the minimum initial ZAMS mass as 5 $M_\odot$. The  and maximum initial mass is the kept to its default value of 150 $M_\odot$. Later, we correct the compact binary yield by multiplying it with the correction factor 0.019.
   

   The redshift-dependent formation efficiency information for the different kinds of compact binaries, viz. NS-NS, NS-BH and BH-BH, together with a metallicity-dependent star formation rate model, is then used to determine the concentration of compact binaries merging at various redshifts following the method described by \cite{2017MNRAS.472.2422M}. We have used a metallicity-independent binary fraction of 0.5 and a redshift limit of $\sim$ 10. Binary fraction is defined as $f_{\rm b}=n_{\rm b}/(n_{\rm s}+n_{\rm b})$ where $n_{\rm b}$ and $n_{\rm s}$ are the number of binary and the number of single star systems, respectively. Thus, a binary fraction of 0.5 implies that one-third of the stars in the ensemble are single whereas two-third of the stars are in binaries.
   
   We chose for our mock data synthesis the model which gave a local merger rate density consistent with the constraint reported in \cite{2023PhRvX..13a1048A}. We adopt an observationally motivated metallicity-dependent star formation rate (SFR), specifically, the model \texttt{206f14SBBoco\_FMR170} from \cite{2021MNRAS.508.4994C} which resembles the Madau-Dickinson SFR \citep{2014ARA&A..52..415M} when summed over metallicities. However, we assert that adopting a perfect binary evolution model or the best metallicity-dependent star formation rate model is not necessary for the problem at hand. The goal is to use an adequately broad intrinsic chirp mass spectrum as an anchor to lift the mass-redshift degeneracy for GW events.

   \subsection{Mock event catalog: Detectability}
   
   The strain (amplitude), $h$, in the interferometer arm of length, $l$, of a GW detector is given by \citep[see, for e.g.][]{2012PhRvD..85l2006A}

   \begin{align}
       h(t)&=-\left(\dfrac{G\mathcal{M}_z}{c^2d_L}\right)\left(\dfrac{t_c-t}{5G\mathcal{M}_z/c^3}\right)^{-1/4} \left(\dfrac{\Theta}{4}\right) \cos{(2\Phi_0+2\Phi_{\rm N})}
   \end{align}
   where $\mathcal{M}=(m_1m_2)^{3/5}/(m_1+m_2)^{1/5}$ is the chirp mass of the binary, and $\mathcal{M}_z=(1+z)\mathcal{M}$ is the redshifted chirp mass. $t_c$ and $\Phi_c$ are time and phase of the binary coalescence i.e. $\Phi(t=t_c)=\Phi_c$. $\Phi_0=\Phi_c+\Phi_0'$ is the termination phase of the GW signal, and
   
   \begin{align}
       \Phi_{\rm N}&=-\left(\dfrac{t_c-t}{5G\mathcal{M}_z/c^3}\right)^{5/8}\\
       \Phi_0^\prime&=-\dfrac{1}{2}\tan^{-1}{\dfrac{2F_\times\cos{\iota}}{F_+(1+\cos^2{\iota})}}
   \end{align}
   where $\iota$ is the orbital inclination of the binary to the line of sight. $\Theta$ is the orientation function of the detector. For the $i$-th detector,

    \begin{align}
    \Theta_i=4\left[F_{+,i}^2\left(\dfrac{1+\cos^2{\iota}}{2}\right)^2+F_{\times,i}^2\cos^2{\iota}\right]^{1/2}
    \end{align}
    such that $0<\Theta_i<4$. $F_{+,i}$ and $F_{\times,i}$ are the antenna pattern functions of the $i$-th interferometer for the $+$ and $\times$ polarizations of the incoming GW signal which are expressed as \citep[see, e.g.][]{2009LRR....12....2S}:

    \begin{align}
    F_{+,1}=\sin\gamma\left[\dfrac{1}{2}(1+\cos^2\theta)\cos 2\phi \cos 2\psi - \cos \theta \sin 2\phi \sin 2\psi\right]\\
    F_{\times,1}=\sin\gamma\left[\dfrac{1}{2}(1+\cos^2\theta)\cos 2\phi \sin 2\psi + \cos \theta \sin 2\phi \cos 2\psi\right]
    \end{align}
    where $\phi$ and $\theta$ are, respectively, the azimuthal and polar angles of the source location in the sky. $\psi$ is the GW polarization angle.
    
    These response functions are smaller by a factor of $\sin \gamma=\sqrt 3/2$ ($\gamma=\pi/3$ for ET-D) compared to those of an L-shaped interferometer with the same arm length. The response functions of the other two interferometers in ET, with arms are obtained from $F_{+,1}$ and $F_{\times,1}$ by the transformation $\phi\to\phi\pm2\pi/3$.

    \vspace{-1em}

    \begin{align}
    F_{+,\times,2}(\theta,\phi,\psi)=F_{+,\times}^{(1)}(\theta,\phi+2\pi/3,\psi)\\
    F_{+,\times,3}(\theta,\phi,\psi)=F_{+,\times}^{(1)}(\theta,\phi-2\pi/3,\psi)
    \end{align}
    
    It can be showed that

    \vspace{-1em}

    \begin{align}
        F_{+,1}^2+F_{+,2}^2+F_{+,3}^2\Big\vert_{\rm max}=3\sin^2\gamma/2=9/8\\
        F_{\times,1}^2+F_{\times,2}^2+F_{\times,3}^2\Big\vert_{\rm max}=3\sin^2\gamma/2=9/8
    \end{align}
    
    Once the redshift distribution of the synthetic compact binary merger population is accomplished, the binaries are then distributed isotropically across the sky and assigned the four angles: ($\phi,\theta,\iota,\psi$). Thereafter, they are checked for detectability using ET's design sensitivity, $S_h(f)$. For this, we calculate the signal-to-noise ratios (SNRs) in ET's three interferometers denoted by $i=1,2,3$.

    \vspace{-1em}
    
    \begin{align}
    \rho_i=2\,\sqrt{\dfrac{5}{24}}\,\left(\dfrac{\Theta_i}{4}\right)\dfrac{(G\mathcal{M}_z)^{5/6}}{\pi^{2/3}c^{3/2}d_L}\sqrt{\displaystyle\int_1^{2f_{\rm max}}\dfrac{df}{f^{7/3}S_h(f)}}
    \end{align}
    \citep[see][]{1996PhRvD..53.2878F,2012PhRvD..86b3502T} where $d_L$ is the luminosity distance of the source, $f$ is the GW frequency, and $2f_{\rm max}$ is the peak GW frequency given by \citep[see, e.g.][]{2012PhRvD..85b3535T}
    
    \vspace{-1em}
    
    \begin{align}
    2f_{\rm max}=\dfrac{c^3}{6\sqrt 6 \pi GM}\approx\dfrac{4400\ \text{Hz}}{1+z}\left(\dfrac{M_\odot}{M}\right)
    \end{align}

    The effective SNR, $\rho_{\rm eff}$, is defined as $\rho_{\rm eff}=\sqrt{\rho_1^2 + \rho_2^2 + \rho_3^2}$
    
    Once the three SNRs for an event are computed, the event is classified as detected or undetected based on the condition: $\rho_i\geq2$ and $\rho_{\rm eff}\geq5$. This lower threshold condition (compared to the commonly used $\rho_i\geq3$ and $\rho_{\rm eff}\geq8$) is chosen to increase the number of detected events which is necessary for cosmography.

    \subsection{Mock event catalog: Observables}

    Once the detectable events are identified, we proceed to determine the observables, $\mathcal{M}_z$ and $d_L$, from the SNRs and phases of the GW signal.

    The redshifted chirp mass, $\mathcal{M}_z$, can be derived from the GW frequency and its time derivative as

    \vspace{-1em}
    
    \begin{align}
    \mathcal{M}_z=\dfrac{c^3}{G}\left(\dfrac{5}{96\pi^{8/3}}\dfrac{\dot f_{\rm GW}}{f_{\rm GW}^{11/3}}\right)^{3/5}
    \end{align}
    where $\dot f_{\rm GW}=df_{\rm GW}/dt$. We assume ${\mathcal{M}_z}$ to be the same as the injected value, albeit, with a Gaussian error of $\mathcal{M}_z/\rho_{\rm eff}$ \citep[see][]{2021PhRvD.104d3014S}. Thus, we get $P({\mathcal{M}_z})$ for all the events.
    
    Furthermore, we assume that the measurement error on each of the three SNRs is $\sigma_\rho=1$, and on each of the three phases is $\sigma_\Phi=\pi/\rho$. We obtain the SNR ratios: $\rho_{21}=\rho_2/\rho_1$ and $\rho_{31}=\rho_3/\rho_1$ and the phase differences: $\Phi_{21}=\Phi_{0,2}-\Phi_{0,1}$ and $\Phi_{31}=\Phi_{3,0}-\Phi_{0,1}$. The probability density of the SNR ratios and the phase differences are given by

    
    \noindent\rule[-9pt]{0.49\textwidth}{0.5pt}\rule[-9pt]{0.5pt}{1em}
    
    \begin{align}
    P(\rho_{j1}) = & \displaystyle\int d\rho_1 \displaystyle\int d\rho_j \,P(\rho_1)\,P(\rho_j)\,\,\delta\big(\rho_{j1} - (\rho_j/\rho_1)\big)\\
    P(\Phi_{j1}) = & \displaystyle\int d\Phi_1 \displaystyle\int d\Phi_j \,P(\Phi_1)\,P(\Phi_j)\,\delta\big(\Phi_{j1} - (\Phi_j - \Phi_1)\big)
    \end{align}
    respectively, where $j=2,3$.\\

    \vspace{-1em}
    
    \begin{align}
    P(\rho_{\rm eff}) = \displaystyle\int d\rho_1 \displaystyle\int d\rho_2 \displaystyle\int d\rho_3\,P(\rho_1)\,P(\rho_2)\,P(\rho_3)\,\delta\big(\rho_{\rm eff} - \rho_0\big)
    \end{align}
    where $\rho_0=\sqrt{\rho_1^2 + \rho_2^2 + \rho_3^2}$ is the measured $\rho_{\rm eff}$.\\

    For convenience, one can use propagation of errors to obtain

    \begin{align}
    \sigma(\rho_{j1})&=\rho_{j1}\sqrt{(1/\rho_j)^2+(1/\rho_1)^2}\\
    \sigma(\Phi_{j1})&=\pi\sqrt{(1/\rho_j)^2+(1/\rho_1)^2}
    \end{align}

    \vspace{-1em}

    \begin{align}
    \sigma(\rho_{\rm eff})&=1
    \end{align}
    and approximate $P(\rho_{21}),P(\rho_{31}),P(\Phi_{21}),(\Phi_{31})$ and $P(\rho_{\rm eff})$ as Gaussians in which case Equations 14--16 are obsolete.

   Following \cite{2021PhRvD.104d3014S}, we outline here the steps to recover the source position in the sky.

   Let $D_1\equiv(\rho_{21},\rho_{31},\Phi_{21},\Phi_{31})$, $\Omega_{\rm sky}\equiv(\phi,\theta)$ and $\Omega_{\rm source}\equiv(\iota,\psi)$ for a given GW event. Then, using the Bayes' theorem,

    \begin{align}
    P(\Omega_{\rm sky},\Omega_{\rm source}|D_1,I)=\dfrac{P(\Omega_{\rm sky},\Omega_{\rm source}|I)P(D_1|\Omega_{\rm sky},\Omega_{\rm source},I)}{P(D_1|I)}
    \end{align}


    As the prior probability $P(\Omega_{\rm sky},\Omega_{\rm source}|I)$ is uniform on both the source and the detector sphere, it becomes:

    
    \begin{align}
    P(\Omega_{\rm sky},\Omega_{\rm source}|I)=P(\Omega_{\rm eff}|I)=\dfrac{1}{(4\pi)^2}
    \end{align}

    
    The likelihood is given by

    \begin{strip}
    \begin{align}
    P(D_1|\Omega_{\rm sky},\Omega_{\rm source},I)&=\displaystyle\int d\rho_{21}P(\rho_{21})\displaystyle\int d\rho_{31}P(\rho_{31})\displaystyle\int d\Phi_{21}P(\Phi_{21})\displaystyle\int d\Phi_{31}P(\Phi_{31})\times \delta(\rho_{21}-\rho_{21}(\Omega_{\rm sky},\Omega_{\rm source})) \nonumber\\
    &\times\ \delta(\rho_{31}-\rho_{31}(\Omega_{\rm sky},\Omega_{\rm source}))\times \delta(\Phi_{21}-\Phi_{21}(\Omega_{\rm sky},\Omega_{\rm source}))\times \delta(\Phi_{31}-\Phi_{31}(\Omega_{\rm sky},\Omega_{\rm source}))
    \end{align}
    \end{strip}
    \noindent where

    \vspace{-1em}

    \begin{align}
        \rho_{j1}(\Omega_{\rm sky},\Omega_{\rm source})&=\Theta_j/\Theta_1=\Theta_{j1}(\Omega_{\rm sky},\Omega_{\rm source})\\[1ex]
        \Phi_{j1}(\Omega_{\rm sky},\Omega_{\rm source})&=\Phi_{0,j}(\Omega_{\rm sky},\Omega_{\rm source})-\Phi_{0,1}(\Omega_{\rm sky},\Omega_{\rm source})
    \end{align}

   
   The posteriors for $\Omega_{\rm sky}$ and $\Omega_{\rm source}$ then become

    \begin{align}
    P(\Omega_{\rm sky}|D_1,I)=\displaystyle\int P(\Omega_{\rm sky},\Omega_{\rm source}|D_1,I)\,d\Omega_{\rm source}\\
    P(\Omega_{\rm source}|D_1,I)=\displaystyle\int P(\Omega_{\rm sky},\Omega_{\rm source}|D_1,I)\,d\Omega_{\rm sky}
    \end{align}

    \vspace{3em}

    \noindent\rule[-2pt]{0.5pt}{1em}\rule[8pt]{0.49\textwidth}{0.5pt}

    The effective orientation function is defined as $\Theta_{\rm eff}=\sqrt{\Theta_1^2+\Theta_2^2+\Theta_3^2}$ so that $0<\Theta_{\rm eff}<6$ (using Equations 4-10). Assuming a flat prior on $\Theta_{\rm eff}$, we have

    \vspace{-1em}
    
    \begin{align}
    P(\Theta_{\rm eff}|I)=\dfrac{1}{\Theta_{\rm eff,\,max}}=\dfrac{1}{6}
    \end{align}

    \vspace{-1em}

    \begin{align}
    P(\Theta_{\rm eff}|D_1,I)=\displaystyle\int d\Omega_{\rm eff}\, P(\Theta_{\rm eff}|I)\,P(\Omega_{\rm eff}|D_1,I)\,\delta(\Theta_{\rm eff}-\Theta_{\rm eff}(\Omega_{\rm eff}))
    \end{align}

    Substituting Equation 20 in 28 gives the $P(\Theta_{\rm eff}$), where the dependence of $\Theta_{\rm eff}$ on the position in the sky, polarization and inclination is explicitly included.

    Lastly, the luminosity distance of the event may be obtained from the expression:
    
    \begin{align}
    P(d_L)=\displaystyle\int d\rho_{\rm eff}P(\rho_{\rm eff})\displaystyle\int d\Theta_{\rm eff}P(\Theta_{\rm eff})\,\delta(d_L-d_L(\rho_{\rm eff},\Theta_{\rm eff}))
    \end{align}
    where $P(\Theta_{\rm eff})=P(\Theta_{\rm eff}|D_1,I)$, and the luminosity distance,

    \begin{align}
    d_L(\rho_{\rm eff},\Theta_{\rm eff})=2\,\sqrt{\dfrac{5}{24}}\,\left(\dfrac{\Theta_{\rm eff}}{4}\right)\dfrac{(G\mathcal{M}_z)^{5/6}}{\pi^{2/3}c^{3/2}\rho_{\rm eff}}\sqrt{\displaystyle\int_1^{2f_{\rm max}}\dfrac{df}{f^{7/3}S_h(f)}}
    \end{align}

This is termed as the recovered luminosity distance. The set of $P(d_L)$ and $P(\mathcal{M}_z)$ pairs constitute the mock GW event catalog. Although impractical, we remove all the NS-NS merger events from the catalog to keep the analysis brief.

   
\subsection{Cosmography}

In the minimal SMOC, the luminosity distance depends on the redshift as

\begin{align}
d_L(z,H_0)=\dfrac{c}{H_0}(1+z)\displaystyle\int_0^{z}\dfrac{dz^\prime}{\sqrt{\Omega_{\rm m}(1+z^\prime)^3+(1-\Omega_{\rm m})}}
\end{align}

The above equation can be numerically inverted to obtain $z(d_L,H_0)$.


In order to do cosmological inference with a broad intrinsic mass spectrum and low SNR-inclusive event catalog, we shall use all the $N$ detected events together in a Monte Carlo fashion. In a given iteration $i$, for every event, we pick a $d_L$ value and an $\mathcal{M}_z$ value based on its $P(d_L)$ and $P(\mathcal{M}_z)$. For a certain event $k$ ($k\in[1,N]$), let us call them $d_{L,\,ik}$ and $\mathcal{M}_{z,\,ik}$. We start with a certain prior of $H_0$, which, in our case, is 60--80 km s$^{-1}$ Mpc$^{-1}$. We divide this range into $J$ equispaced $H_0$ values. In our simulation, we use $J=200$. The narrow prior for $H_0$ has been used to reduce the computation time. One may use a wider prior for $H_0$ as well and the results will be similar. For each $H_0$ value (indexed by $j$), we find the corresponding $z_{ijk}$ value using $z_{ijk} = z(d_{L,ik},H_{0,j})$, and then the $\mathcal{M}_{ijk}$ value using $\mathcal{M}_{ijk}=\mathcal{M}_{z,ik} / (1+z_{ijk})$.



   \begin{figure}[!ht]
   \centering
   \includegraphics[width=0.48\textwidth]{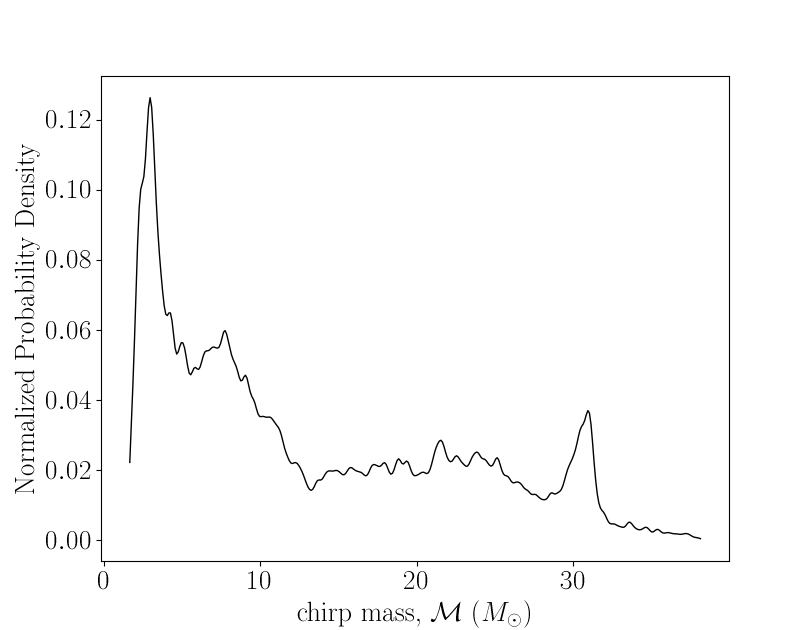}
   \caption{Intrinsic chirp mass distribution of the merging NS-BH and BH-BH systems. The chirp mass range of the injected sources is 1.6--40 $M_\odot$.}
   \label{chm_hist}
   \end{figure}

   \begin{figure*}
   \centering
   \includegraphics[width=0.48\textwidth]{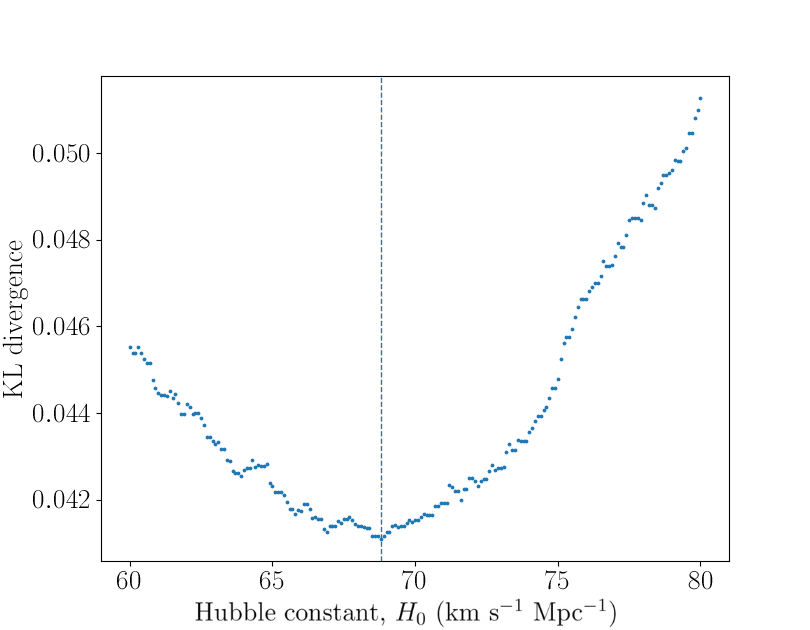}
   \includegraphics[width=0.48\textwidth]{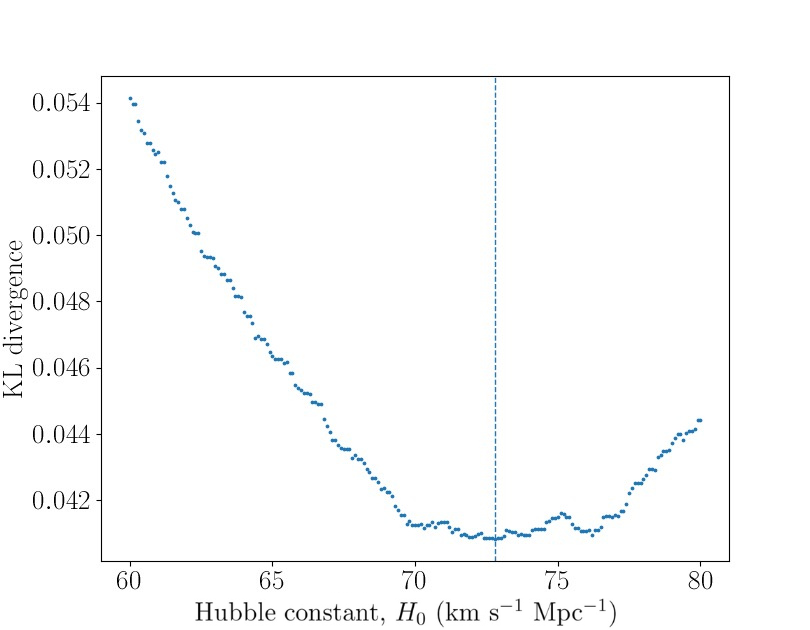}
   \caption{Scatter plot of KL divergences for a certain iteration for Case I (\emph{left}), and for a certain iteration for Case II (\emph{right}) with the dashed line showing the minimum. The $H_0$ value corresponding to this is taken to be the best $H_0$ for these iterations.}
   \label{h_best}
   \end{figure*}

   \begin{figure*}
   \centering
   \includegraphics[width=0.9\textwidth]{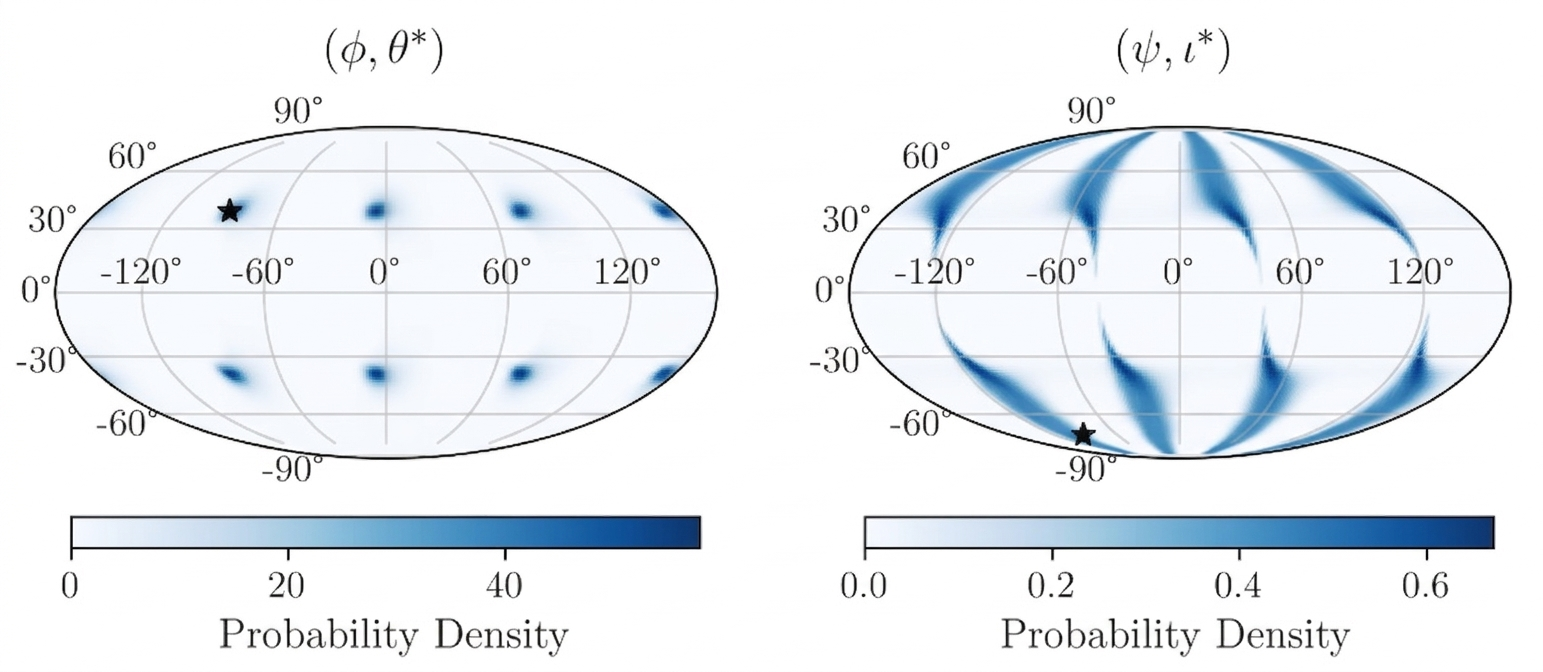}
   \caption{Localization for $(\phi,\theta^*)$ and $(\psi,\iota^*)$ recovered for a certain event with $\rho_{\rm eff}=82.66$. The coordinates, $\theta^*=90\degree-\theta$ and $\iota^*=90\degree-\iota$ are used instead of $\theta$ and $\iota$, respectively. The stars denote the injected coordinates of the source.}
   \label{localization}
   \end{figure*}
   
   \begin{figure*}
   \centering
   \includegraphics[width=0.48\textwidth]{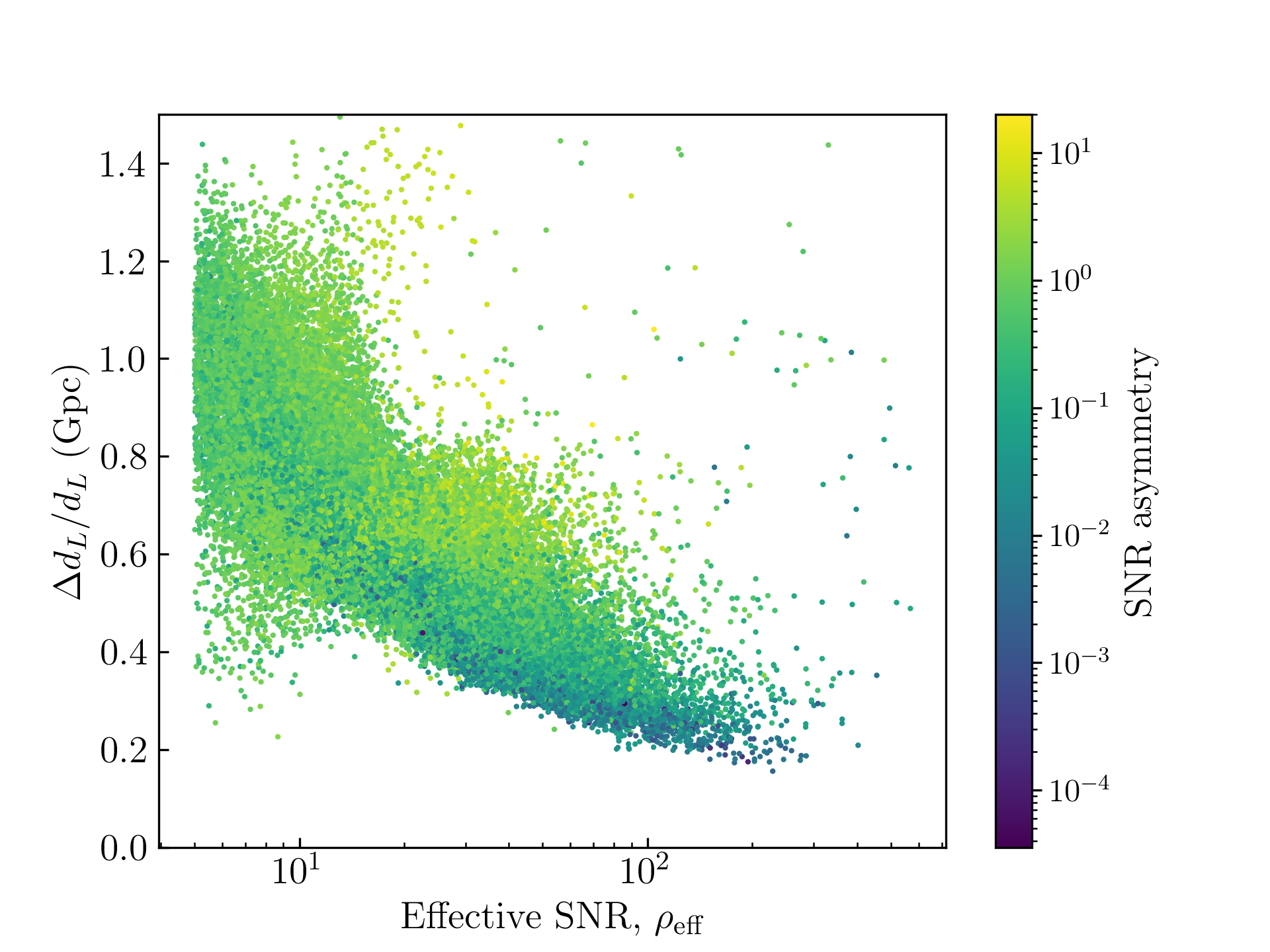}
   \includegraphics[width=0.48\textwidth]{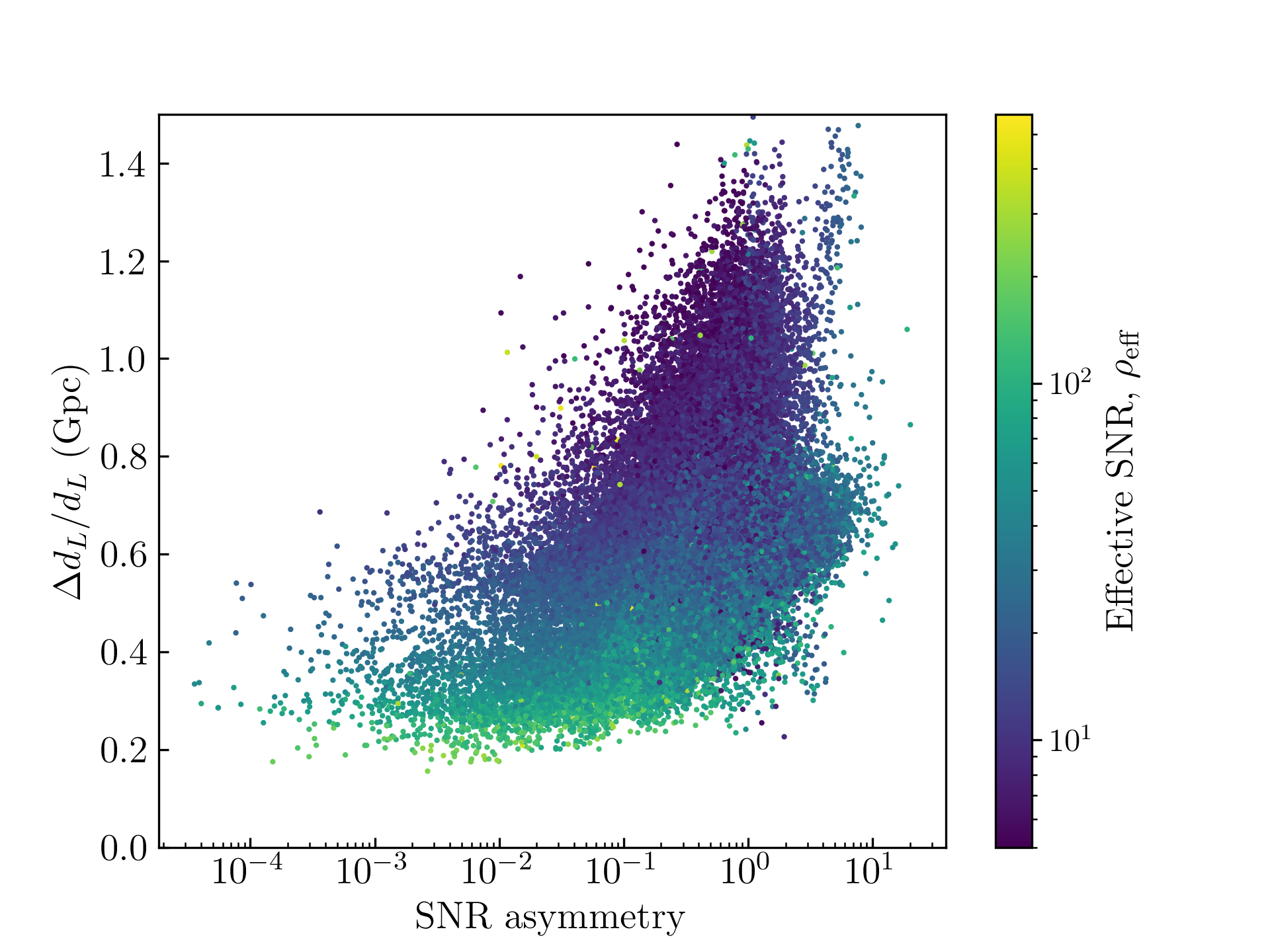}
   \caption{Plot of relative error $d_L$ versus effective SNR (\emph{left}), and relative error $d_L$ versus SNR asymmetry (\emph{right}), for dataset 1. Both effective SNR and SNR asymmetry decide the accuracy of the recovered $d_L$.}
   \label{rel_err}
   \end{figure*}
   
   \begin{figure*}
   \centering
   \includegraphics[width=0.48\textwidth]{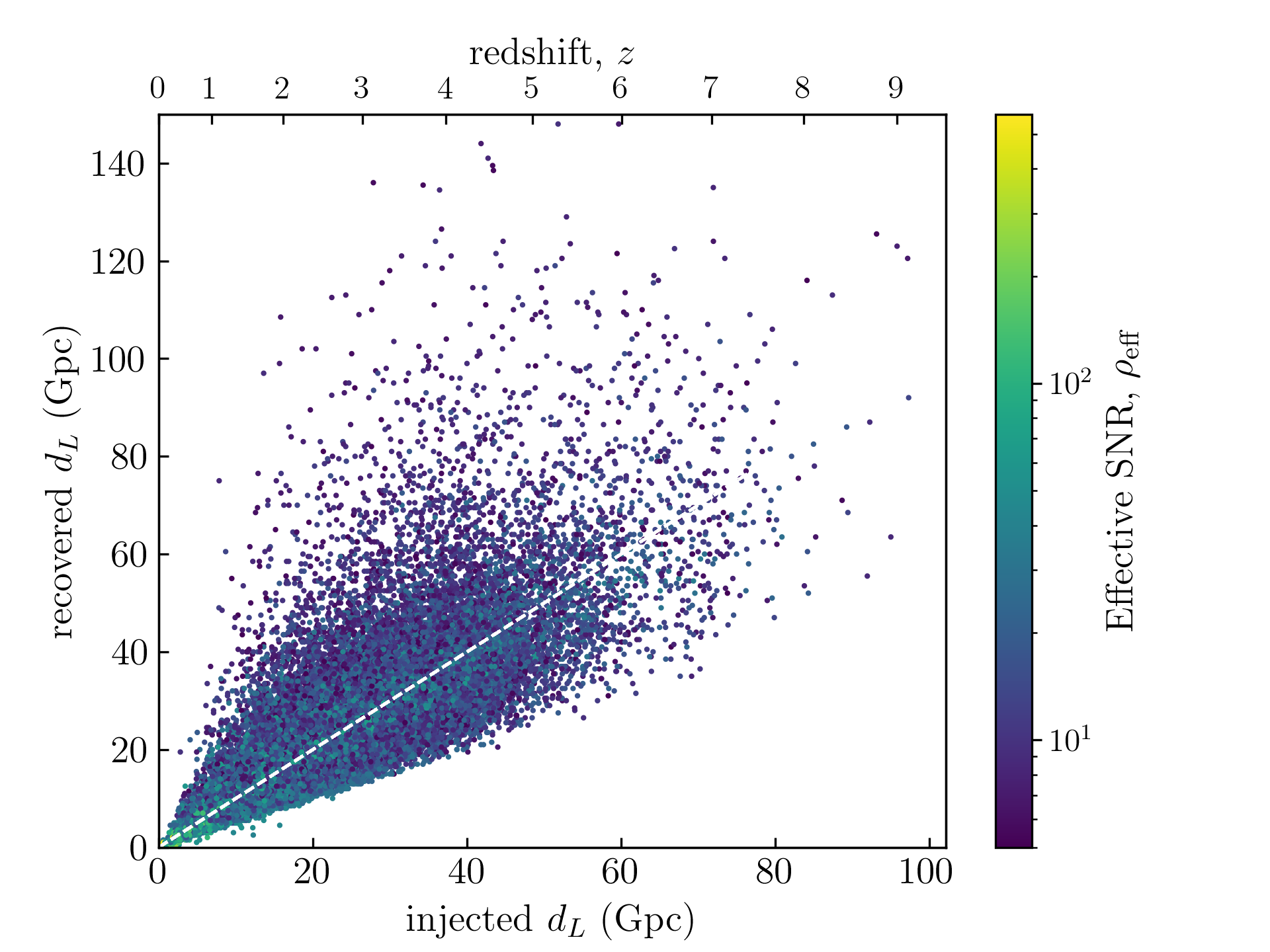}
   \includegraphics[width=0.48\textwidth]{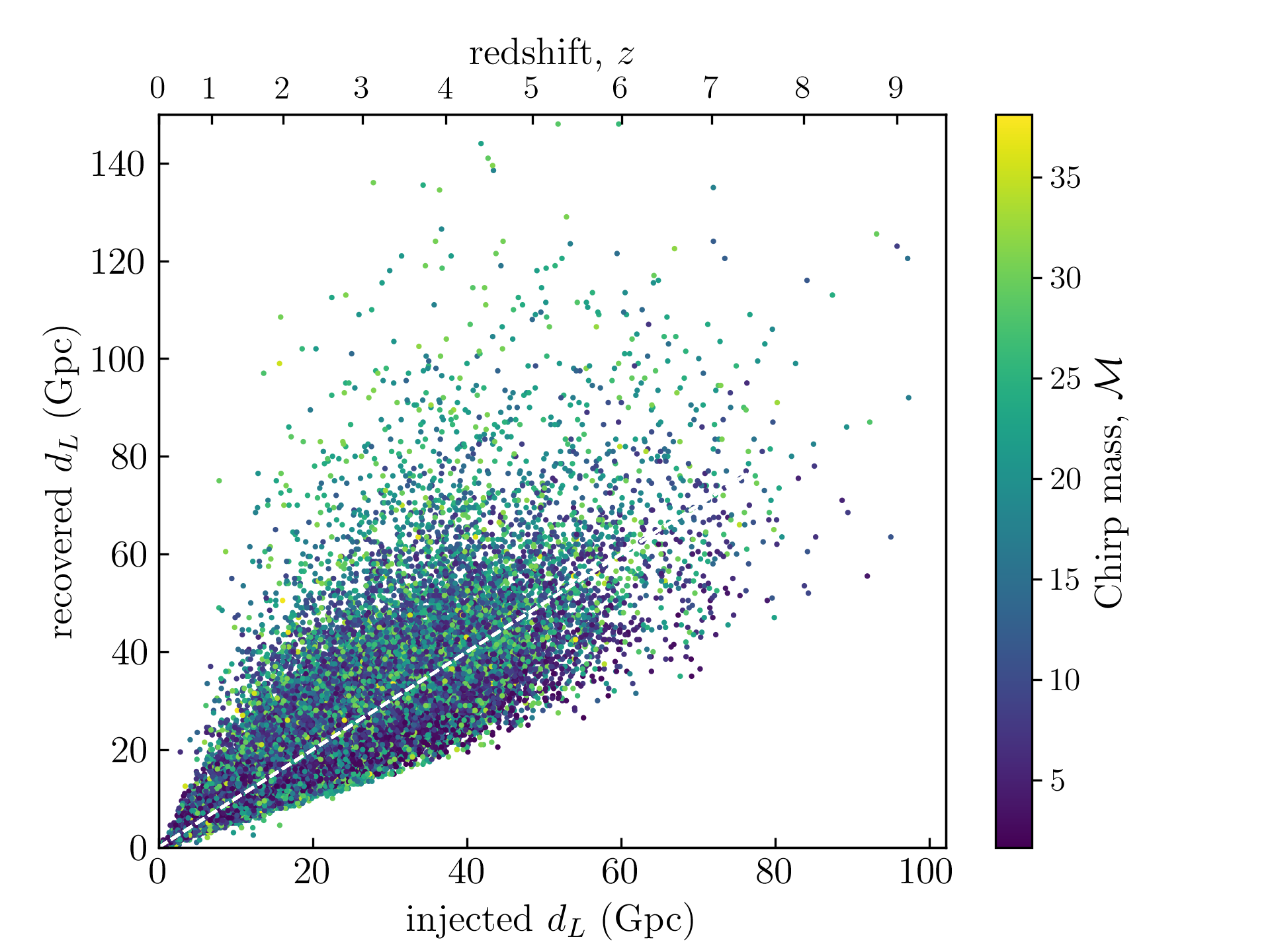}
   \caption{Plot of recovered $d_L$ versus injected $d_L$ for dataset 1 to check the sanity of mock data. The white dashed line corresponds to $45\degree$ slope representing ideal recovery. No evident bias is seen.}
   \label{sanity}
   \end{figure*}

We then collect the $\mathcal{M}_{ijk}$ for all the events and obtain the probability distribution, $P_{ij}(\mathcal{M})$, and compare it with the model (expected) intrinsic distribution, $P(\mathcal{M})$ (Figure~\ref{chm_hist}), by finding the Kullback–Leibler (KL) divergence, $D_{\text{KL},ij}\left(P(\mathcal{M})\parallel P_{ij}(\mathcal{M})\right)$.

   \begin{align}
   P_{ij}(\mathcal{M})\approx \sum_l{C_l\,\text{rect}\left(\dfrac{\mathcal{M}_{ijk}-\mathcal{M}_l}{W}\right)}
   \end{align}
   where $C_l$ is the count in the $l$-th chirp mass bin, and $W$ is the binwidth. The chirp mass distribution is divided into 100 bins, so $l\in[1,100]$. ``rect" refers to the rectangle function defined by

   \begin{align}
       \text{rect}\,(x)=\begin{cases}1, & |x|\leq 1/2\\0, & \text{otherwise}\end{cases}
   \end{align}

   It is important to use a chirp mass range much broader than the intrinsic chirp mass range in order to account for the errors on $d_L$ and $\mathcal{M}_z$ which may yield $\mathcal{M}_{ijk}$ outside the intrinsic chirp mass range. We use the chirp mass range $[0.1,150]\,M_\odot$.
   
   \begin{figure*}
   \centering
   \includegraphics[width=0.48\textwidth]{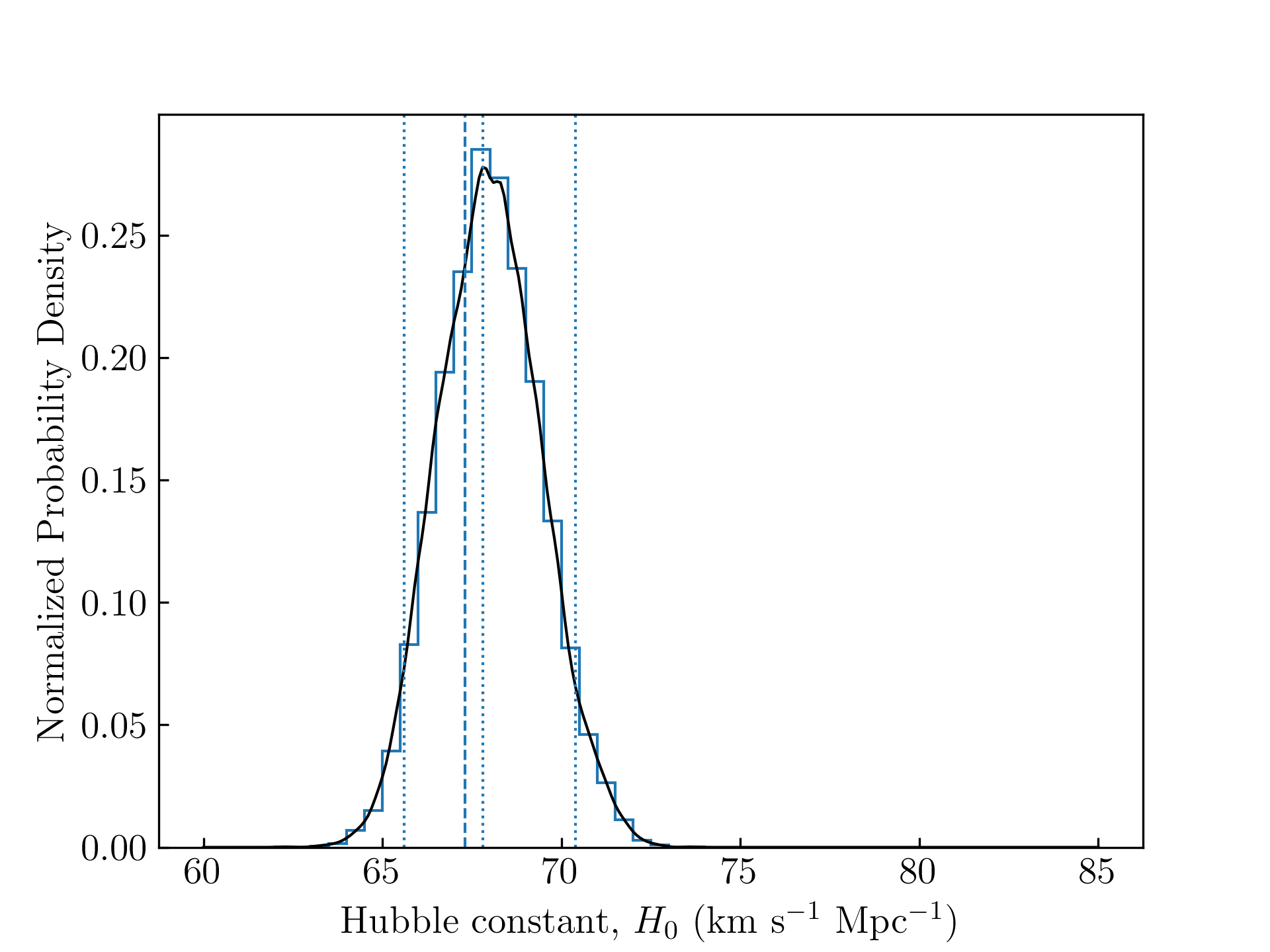}
   \includegraphics[width=0.48\textwidth]{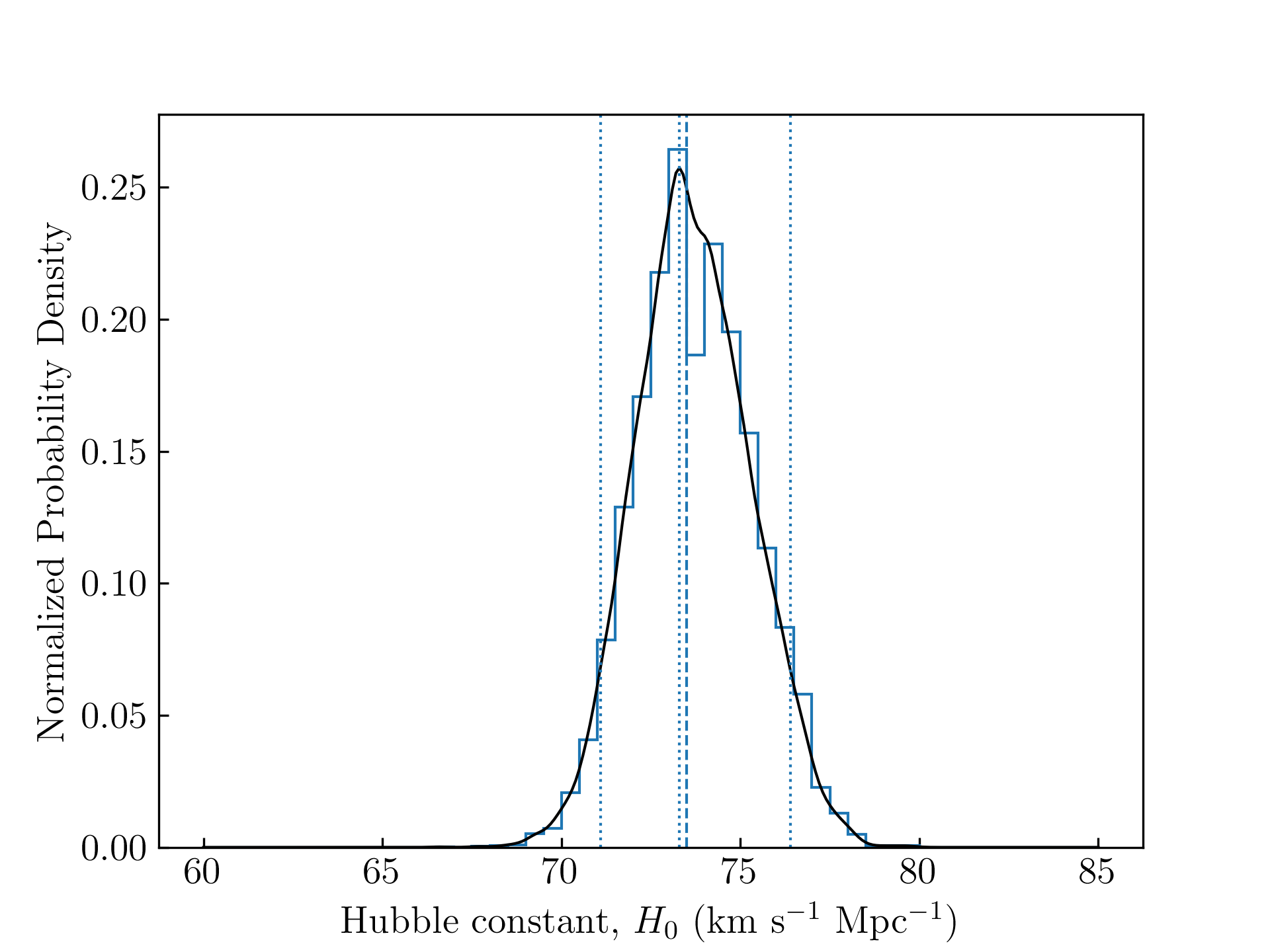}
   \caption{Histograms of the $H_0$ values obtained from 10000 iterations for case I (\emph{left}) and for case II ((\emph{right}). The binsize is 0.5 km s$^{-1}$ Mpc$^{-1}$. The dashed line shows the injected value. The middle dotted line shows the mode of the distribution. The left and the right dotted lines show the lower and upper bounds, respectively. The smooth curve is the kernel density estimate.}
   \label{h_hist}
   \end{figure*}

   \begin{figure}[!ht]
   \centering
   \includegraphics[width=0.48\textwidth]{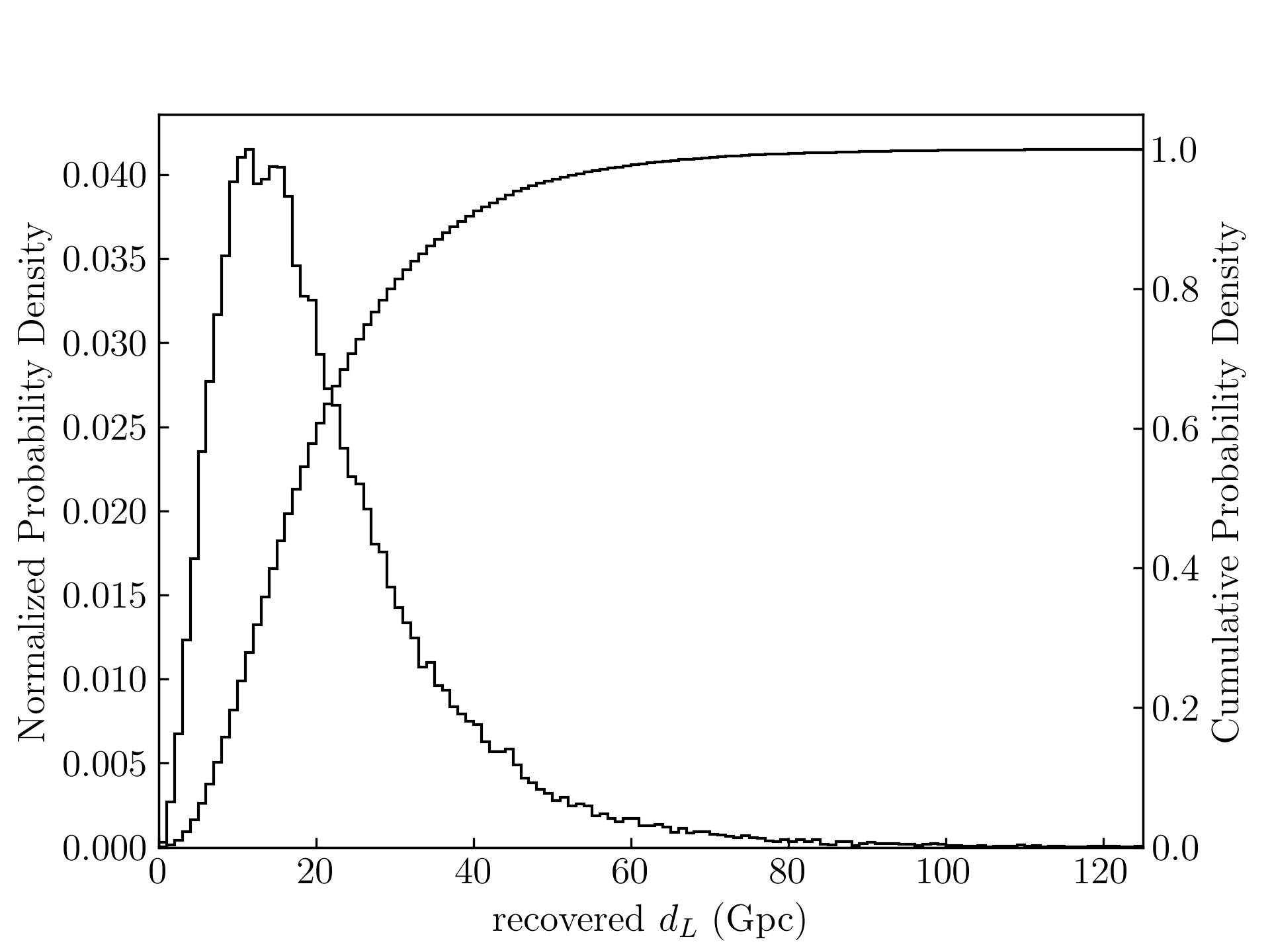}
   \caption{Distribution and cumulative distribution of recovered luminosity distance of all the NS-BH and BH-BH sources in dataset 1.}
   \label{pop_hist}
   \end{figure}

   \begin{figure*}
   \centering
   \includegraphics[width=0.48\textwidth]{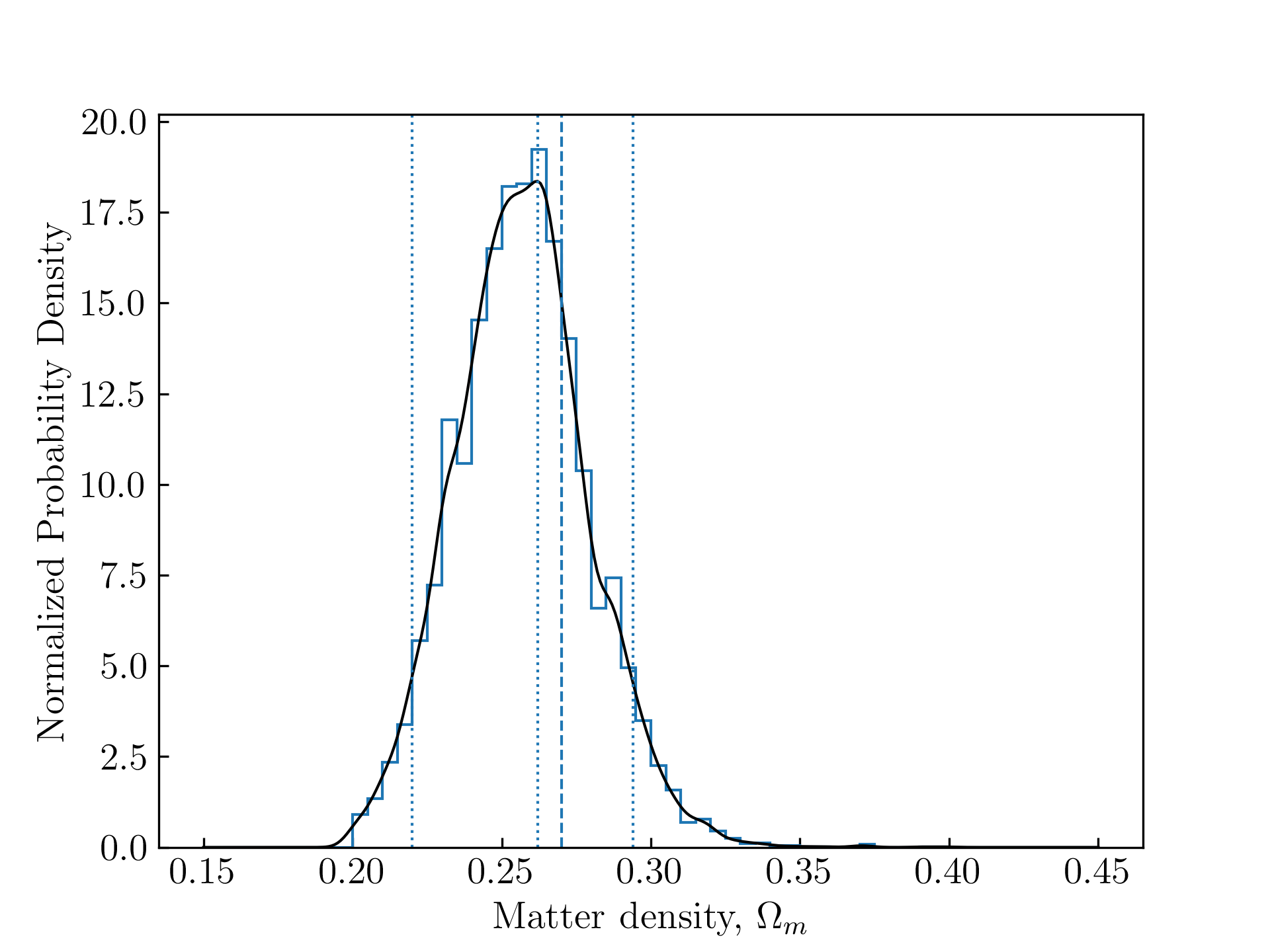}
   \includegraphics[width=0.48\textwidth]{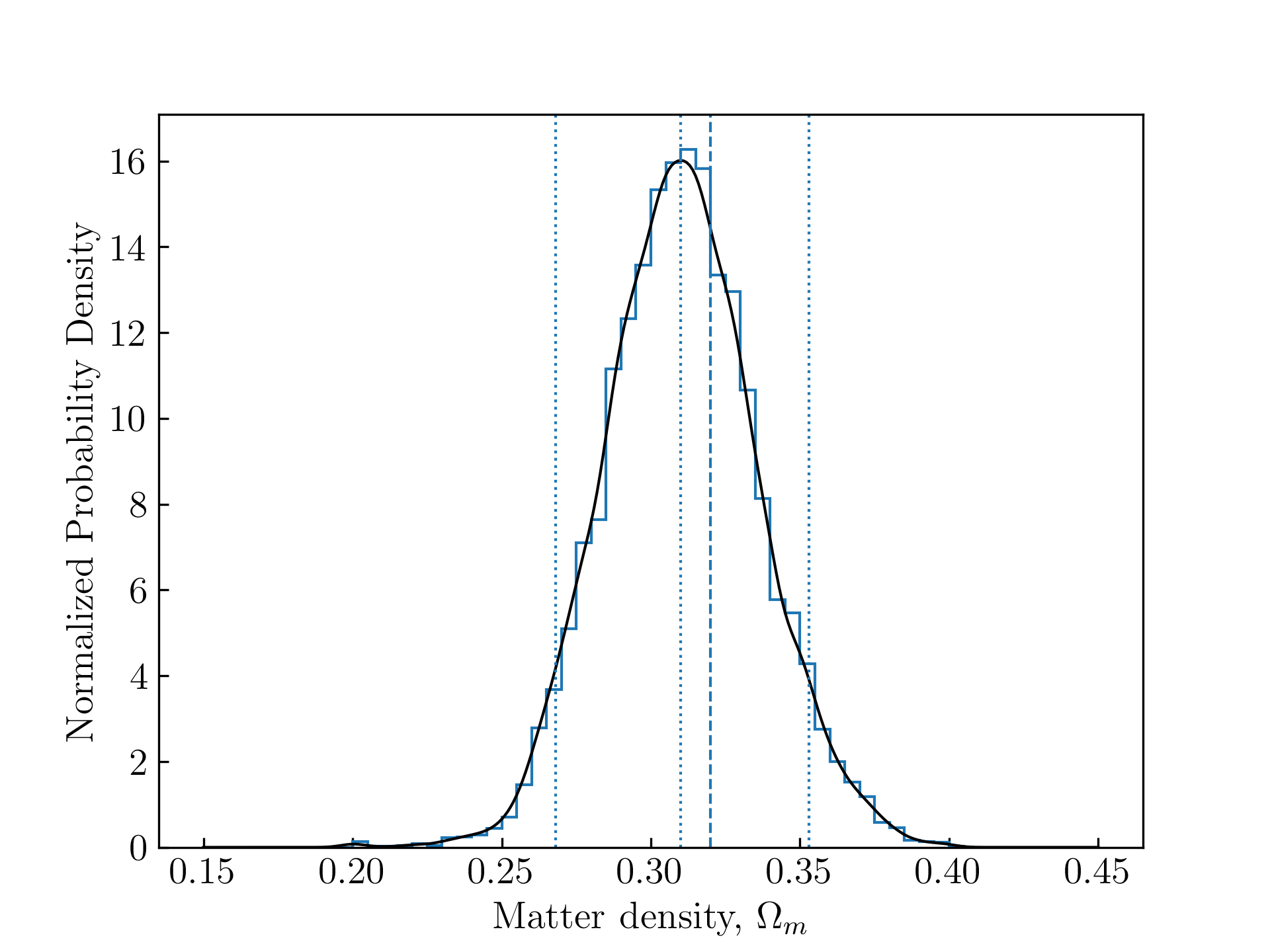}
   \caption{Histograms of the $\Omega_{\rm m}$ values obtained from 10000 iterations for case III (\emph{left}) and for case IV ((\emph{right}). The binsize is 0.005. The dashed line shows the injected value. The middle dotted line shows the mode of the distribution. The left and the right dotted lines show the lower and upper bounds, respectively. The smooth curve is the kernel density estimate.}
   \label{om_hist}
   \end{figure*}

   \begin{figure*}
   \centering
   \includegraphics[width=0.48\textwidth]{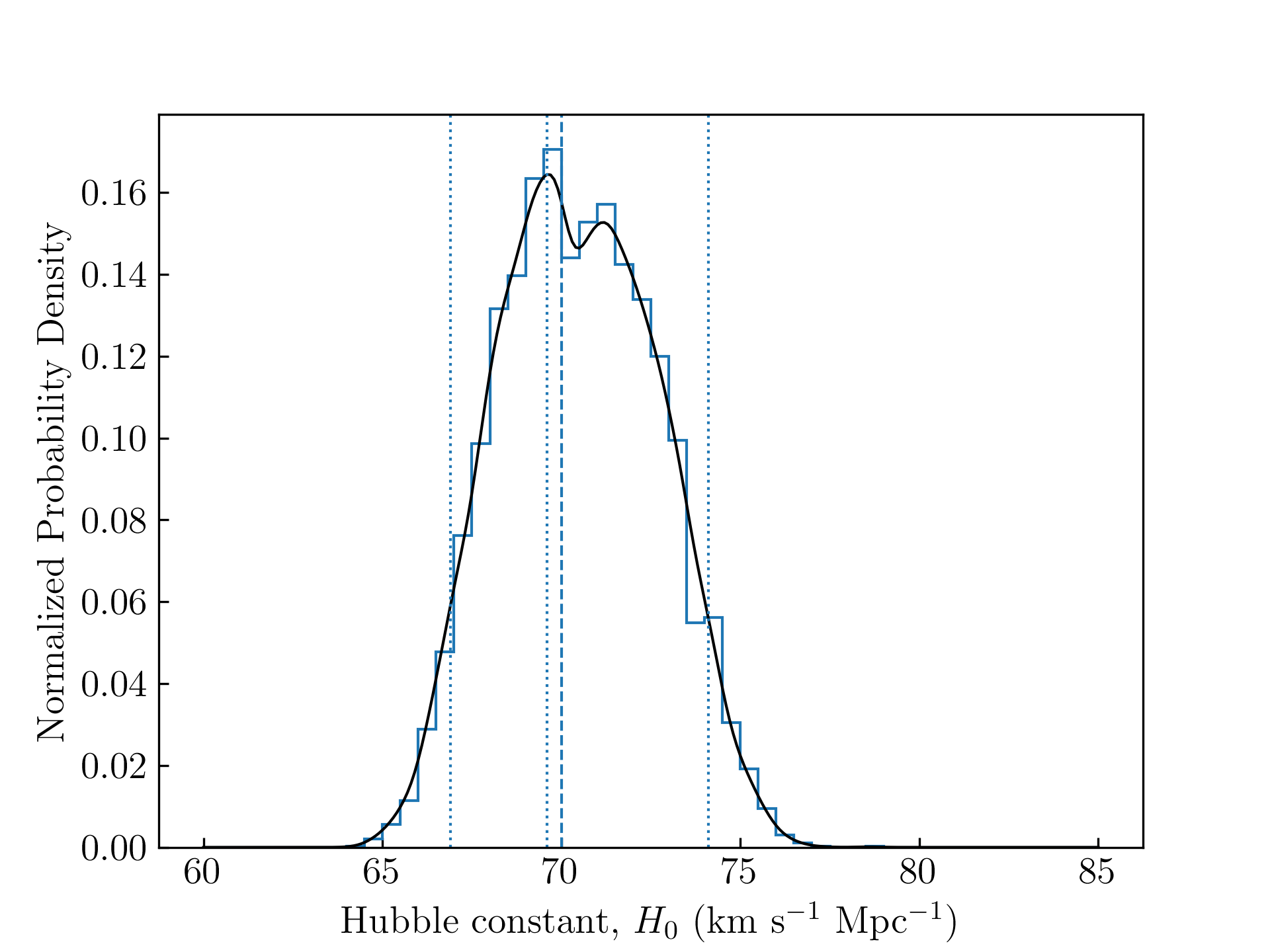}
   \includegraphics[width=0.48\textwidth]{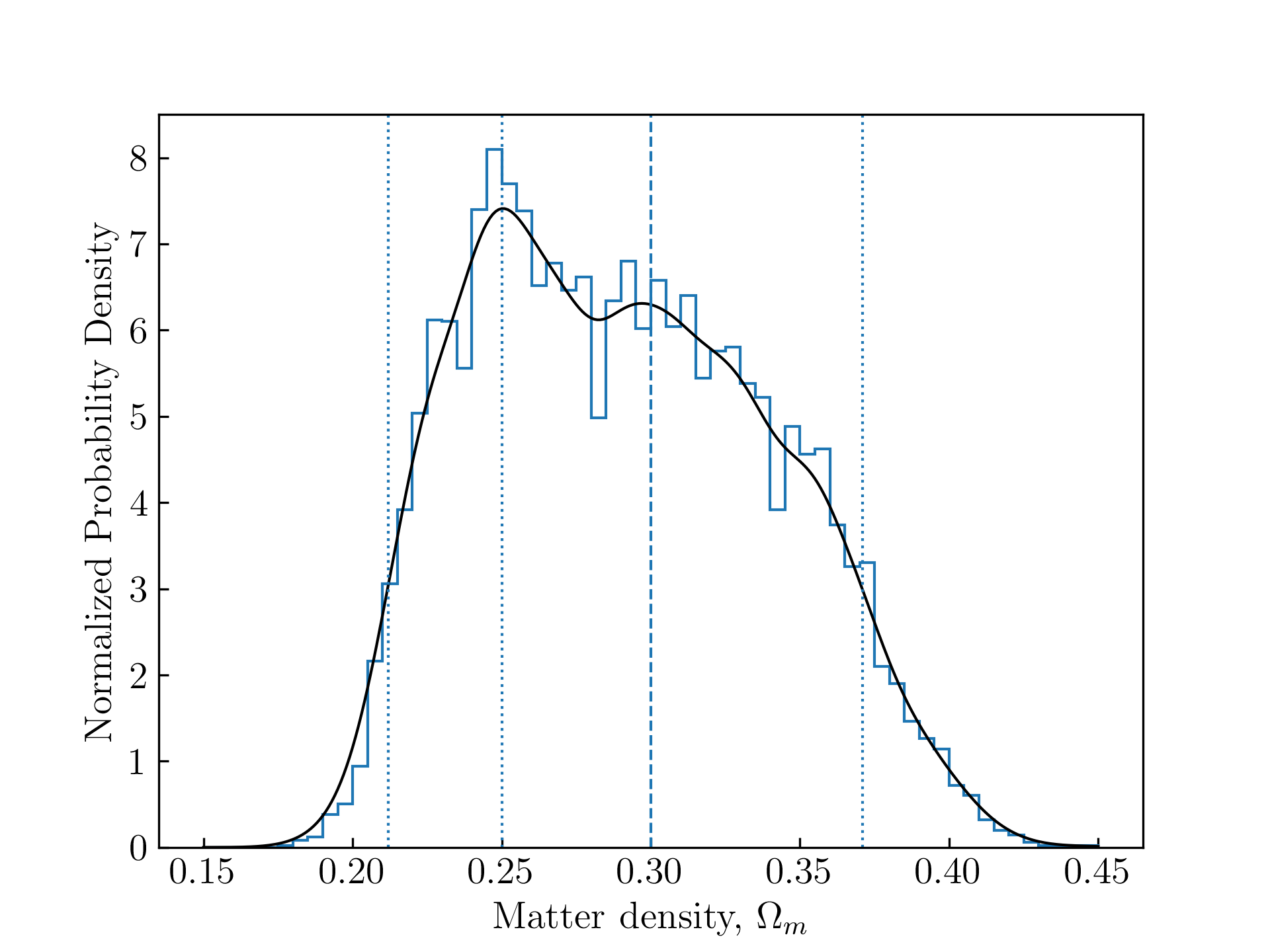}
   \caption{Histograms of the $H_0$ values for case IX (\emph{left}) and of the $\Omega_{\rm m}$ values (\emph{right}) for case X (\emph{right}) obtained from 10000 iterations using dataset 5+6. The binsize is 0.005. The dashed line shows the injected value. The middle dotted line shows the mode of the distribution. The left and the right dotted lines show the lower and upper bounds, respectively. The smooth curve is the kernel density estimate.}
   \label{h_om_hist}
   \end{figure*}

   \begin{figure*}
   \centering
   \includegraphics[width=0.48\textwidth]{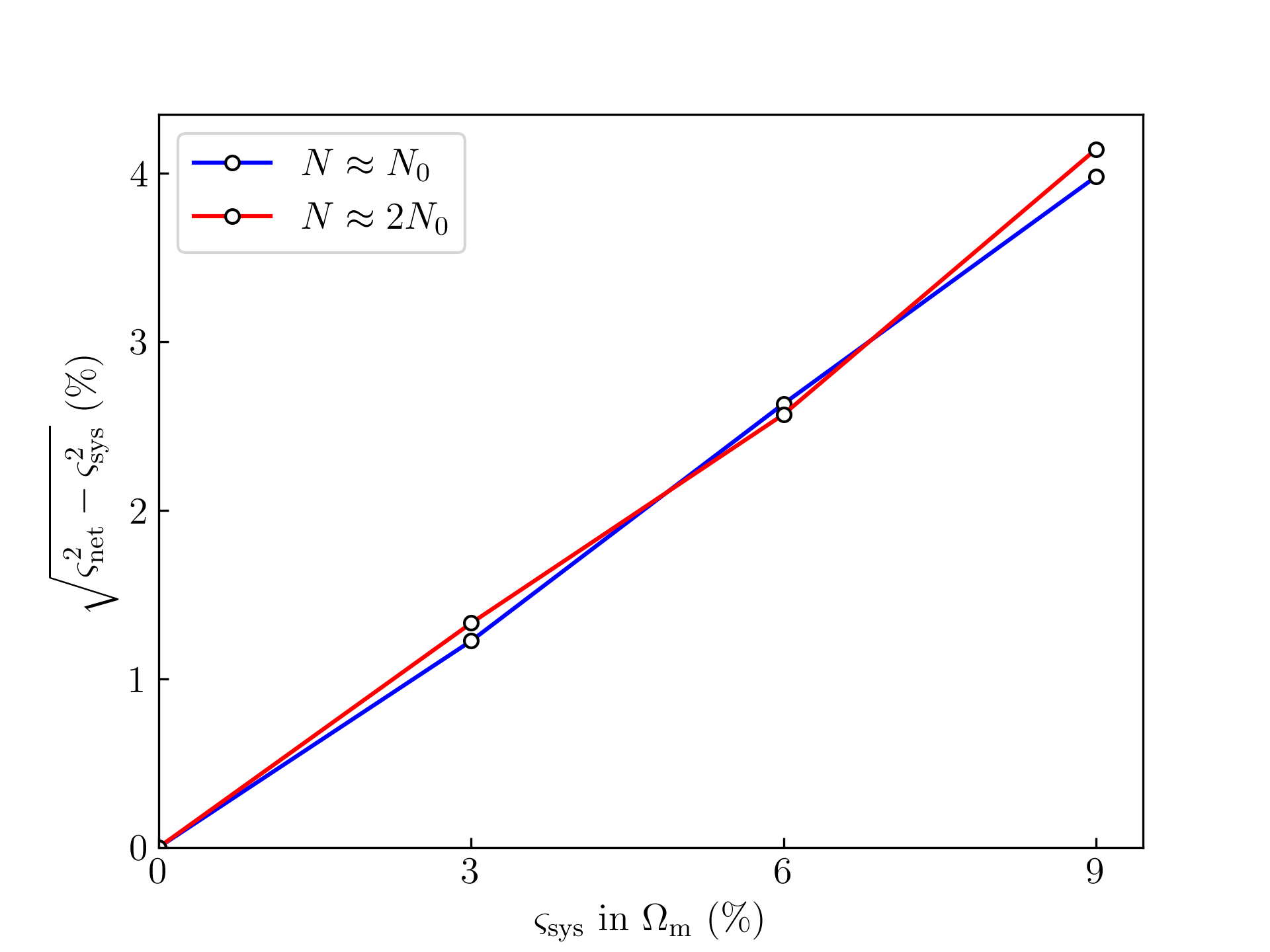}
   \includegraphics[width=0.48\textwidth]{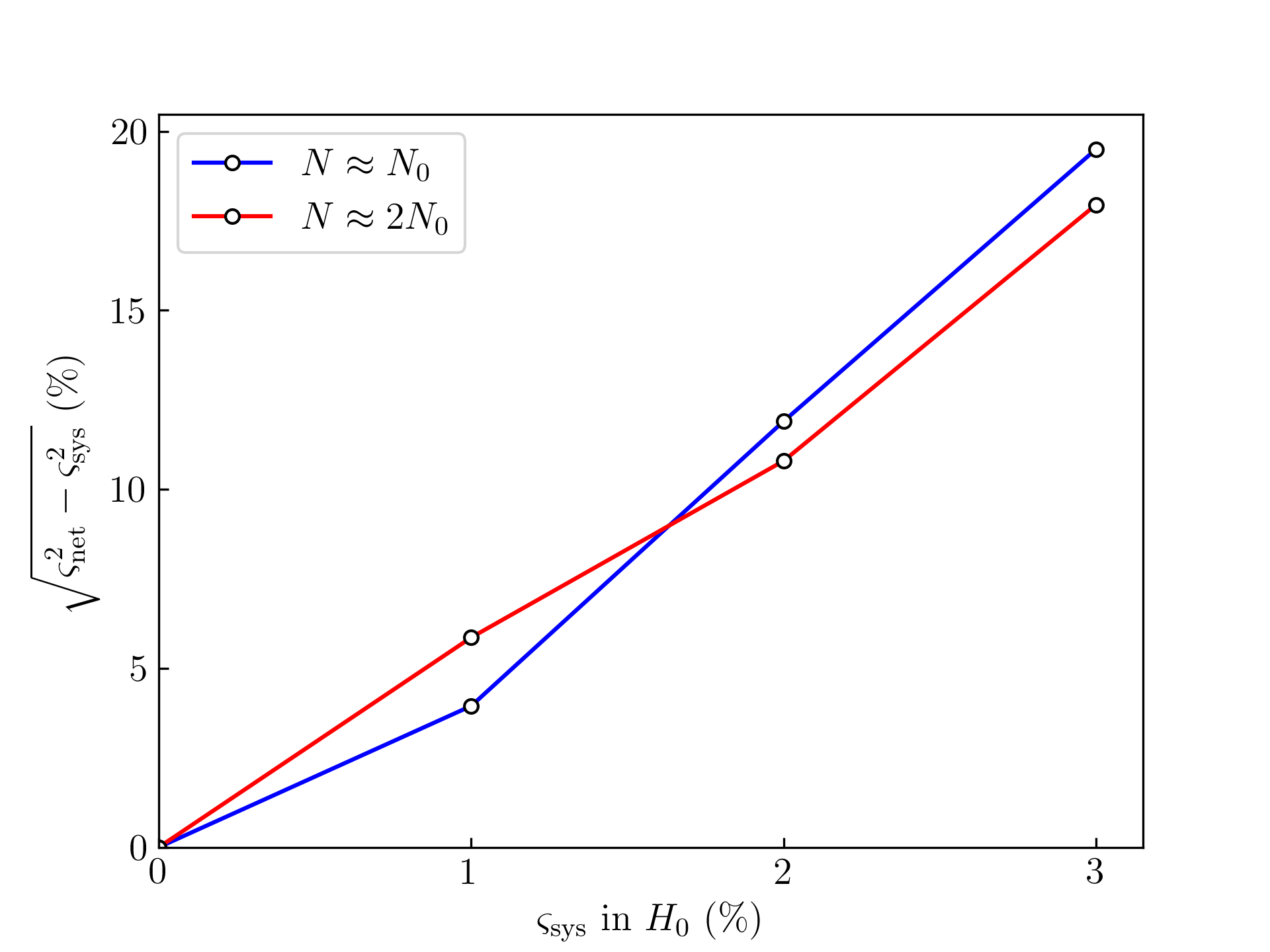}
   \caption{Plot of $\sqrt{\varsigma_{\rm net}^2-\varsigma_{\rm stat}^2}$ versus the systematic uncertainty, $\varsigma_{\rm sys}$, for $H_0$ (\emph{left}) and for $\Omega_{\rm m}$ (\emph{right}). The values for $N \approx N_0$ are the averages of the values for datasets 5 and 6. The values for $N \approx 2N_0$ are the values for dataset 5+6.}
   \label{h_om_err}
   \end{figure*}

If $P$ is the true probability distribution and $Q$ is an approximating probability distribution, the KL divergence \citep{Kullback_1951}, also known as relative entropy, is defined as

\begin{align}
D_{\text{KL}}(P\parallel Q)=\sum_{x\in\mathcal{X}} P(x)\ln\dfrac{P(x)}{Q(x)}    
\end{align}
where $\mathcal{X}$ is the set of bins into which both the distributions are divided, say 100. $P(x)$ and $Q(x)$ are the probability density values at $x$. We use KL divergence against other divergences because it is directional and asymmetric as evident from its definition.


Out of the $J$ $D_{\text{KL},ij}$ values, the best $H_0$ for the $i$-th iteration is the one for which $D_{\text{KL},ij}$ is minimum (see Figure~\ref{h_best}).

\begin{align}
    H_{0,i}=H_{0,j}:D_{\text{KL},i}=\text{min}_j(D_{\text{KL},ij})
\end{align}
We perform $n=10000$ such iterations and create a histogram of the $H_0$ values. From the kernel density estimate of the $H_0$ data, we derive the global best $H_0$. Therefore, we may express the probability distribution as


\begin{align}
   \hat{P}(H_0)=\dfrac{1}{nh}\sum_{i=1}^n K\left(\dfrac{H_0-H_{0,i}}{h}\right)
\end{align}
\citep[see][]{hastie01statisticallearning} where the bandwidth, $h=n^{-0.2}\approx0.16$ by Scott's rule \citep{scott1992multivariate}. $K(x)$ is a Kernel function.





For the Gaussian kernel, the above equation becomes

\begin{align}
\hat{P}(H_0)=\dfrac{1}{n}\sum_{i=1}^n \dfrac{1}{(h\sigma)\sqrt{2\pi}}\exp{\left(\dfrac{-(H_0-H_{0,i})^2}{2(h\sigma)^2}\right)}
\end{align}
where $\sigma$ is the sample standard deviation.

Next, we carry out a similar analysis to recover $\Omega_{\rm m}$, the prior for which is chosen to be 0.2--0.4.

\section{Results}

Once an event in the set of injected GW events is classified as detected based on the SNR threshold, it is taken up for localization and subsequently, distance estimation. An example of source localization is shown in Figure~\ref{localization}. The multiple zones in the maps are due to the various angular symmetries in the Equations 5--8. Following Equations 29--31, one gets the probable distance i.e. $P(d_L)$ of the source whose accuracy depends both on the effective SNR and the SNR ratios (see Figure~\ref{rel_err}). The higher the effective SNR and the farther the SNR ratios from unity, the better the recovery of the source position. While the effective SNR is primarily effected by the chirp mass and the source distance, the SNR ratios are effected by the coordinates of the source.


The SNR asymmetry, $Y(\rho)$, may be defined as

\vspace{-1em}

\begin{align}
    Y(\rho)=\text{max}(\rho_j/\rho_i,\rho_i/\rho_j)-1
\end{align}
where $i,j\in[1,2,3]$ and $i\neq j$. The $\max$ function ensures that the ratio is greater than unity. For perfect symmetry i.e. $\rho_1=\rho_2=\rho_3$, $Y(\rho)=0$.

We prepare six mock datasets. Each dataset consists of the detected BH-BH and BH-NS  events from the $10^5$ injected binary merger events in some standard cosmological model characterized by $H_0$ and $\Omega_{\rm m}$. With the detection criteria as $\rho_{\rm eff}\geq5$, we detect around 55000 BH-NS and BH-BH events in all datasets, leaving out the NS-NS events for simplicity. In every dataset, around 5/6-th of the injected BH-BH and BH-NS events are detected with ${\rm SNR}\geq5$. The plot of recovered $d_L$ versus injected $d_L$ for one of the datasets is shown in Figure~\ref{sanity}. The injected values of the ($H_0, \Omega_m$) pair for each case are detailed in Table~\ref{overview}.
   
We use the binary BH intrinsic chirp mass spectrum derived from simulation. In the first case, with 54850 detected mock BH-BH and BH-NS events, we recover an $H_0$ value of $67.8_{-2.2}^{+2.6}$ km s$^{-1}$ Mpc$^{-1}$ when the injected value is $67.3$ km s$^{-1}$ Mpc$^{-1}$. Whereas, in the second case, with 56455 detected events, we get an $H_0$ value of $73.3_{-2.2}^{+3.2}$ km s$^{-1}$ Mpc$^{-1}$ when the injected value is $73.5$ km s$^{-1}$ Mpc$^{-1}$ (see Figure~\ref{h_hist}). The quoted errors correspond to the 90\% credible interval. Thus, with $\sim\,5\times10^4$ events, the statistical uncertainty in the inferred $H_0$ is around 3.5\%. Cases I and II show that the recovered $H_0$ tracks the injected $H_0$.

It may be noted that $\sim$ 90\% of the detected mergers originate from $d_L<$ 40 Gpc i.e. $z<\,\sim4$ (see Figure~\ref{pop_hist}). So, mergers from higher redshifts do not have significant contribution to the $H_0$ or the $\Omega_{\rm m}$ estimate.

We then proceed to infer $\Omega_{\rm m}$ keeping $H_0$ fixed for two cosmological cases, III and IV. With 55105 detected binary BH events, for case III, we recover an $\Omega_{\rm m}$ value of $0.262_{-0.042}^{+0.032}$ against an injected value of 0.27. For case IV, we recover an $\Omega_{\rm m}$ value of $0.310_{-0.042}^{+0.043}$ against an injected value of 0.32 with 55767 events (see Figure~\ref{om_hist}). Cases III and IV show that the recovered $\Omega_{\rm m}$ tracks the injected $\Omega_{\rm m}$.

Cases V and VI are similar to case I / II and case III / IV, respectively, but with new datasets. Next, we consider six cases wherein a systematic Gaussian uncertainty is assumed in the value of the known parameter. Lastly, we take up two cases, XIII and XIV, wherein the known parameter is specified to lie in a small range with a uniform distribution. For each of these eight cases, we use three datasets of which the third one is the combination of the other two datasets. The third dataset was used to see if the statistical uncertainty in the parameter estimation drops in the same manner (i.e. by $1 / \sqrt{N}$) as in the cases V and VI where the known parameter has no systematic uncertainty. All these ten cases have been explored with mock datasets 5 and 6 both of which concern the cosmological model: $[H_0,\Omega_{\rm m}]=[70,0.3]$.

For cases VII, IX and XI, we assume that $\Omega_{\rm m}$ is known with an uncertainty of 3\%, 6\% and 9\%, respectively. This results in 3.1\%, 3.8\% and 5\% uncertainties in the $H_0$ estimate with 111009 events. With roughly half the number of events, the uncertainties are $\sim$3.9\%, $\sim$ 4.5\% and $\sim$5.4\%, respectively. For cases VIII, X and XII, we assume that $H_0$ is known with an uncertainty of 1\%, 2\% and 3\%, respectively. This results in 12.2\%, 15.2\% and 20.9\% uncertainties in the $\Omega_{\rm m}$ estimate with 111009 events. With roughly half the number of events, the uncertainties are $\sim$14.5\%, $\sim$18.3\% and $\sim$24.9\%, respectively. Thus, the decline in the statistical uncertainty deviates from the $1 / \sqrt{N}$ rule as the systematic uncertainty in the known parameter increases.


In case XIII, we assume that $\Omega_{\rm m}$ can be anywhere in the range $[0.25,0.35]$ with a uniform probability. Using 111009 events, we recover an $H_0$ value of $69.6_{-2.7}^{+4.5}$ km s$^{-1}$ Mpc$^{-1}$ (see Figure~\ref{h_om_hist}) and the uncertainty is 5.2\%. Likewise, in case XIV, we assume that $H_0$ can be anywhere in the range $[65,75]$ km s$^{-1}$ Mpc$^{-1}$ with a uniform probability. Using 111009 events, we recover an $\Omega_{\rm m}$ value of $0.250_{-0.038}^{+0.121}$ (see Figure~\ref{h_om_hist}) and the uncertainty is 31.8\%. Compared to when only about half of the events are considered, the relative decline in the uncertainty is only $\sim$10\% for cases XIII and XIV.

The results have been summarized in Table~\ref{overview}. The results for dataset 5+6 are obtained from separate runs, not by combining the results for datasets 5 and 6.

In the presence of a systematic uncertainty on the known parameter, the net uncertainty takes the form:


\begin{align}
    \varsigma_{\rm net}=\sqrt{\varsigma_{\rm stat,0}^2/\lambda+\kappa^2\varsigma_{\rm sys}^2}
\end{align}
where $\varsigma$ denotes the relative uncertainty with $N$ events, $\varsigma_{\rm stat,0}$ is the net uncertainty with $N_0$ events when the systematic uncertainty is absent, and $\kappa$ is a scaling factor. $N_0 \approx 55000$ is the reference event count, and $\lambda=N/N_0$ such that $\varsigma_{\rm net}=\varsigma_{\rm stat,0}/\sqrt{\lambda}=\varsigma_{\rm stat}$ when $\varsigma_{\rm sys}=0$. 

We plot $\sqrt{\varsigma_{\rm net}^2-\varsigma_{\rm stat}^2}$ listed in Table~\ref{overview} against the systematic uncertainties for $H_0$ and also, for $\Omega_{\rm m}$, and find the relation to be linear for both i.e. the factor, $\kappa$, does not depend on $\varsigma_{\rm sys}$ (see Figure~\ref{h_om_err}). From the best fit using the method of least squares and setting the $y$-intercept to be zero, we get $\kappa \approx 0.439$ for $H_0$, and $\kappa \approx 6.162$ for $\Omega_{\rm m}$ with $N \approx N_0$, and $\kappa \approx 0.450$ for $H_0$, and $\kappa \approx 5.808$ for $\Omega_{\rm m}$ with $N \approx 2N_0$. $\kappa$ appears to be independent of $N$ but it has not been verified due to computational limitations. Whether $\kappa$ depends on the value of the known parameter also remains to be ascertained.

Using Equation 36, and assuming $\kappa \approx 0.445$ for $H_0$ and $\kappa \approx 5.985$ for $\Omega_{\rm m}$, one can estimate the net uncertainty for any number of events and for arbitrary value of systematic uncertainty from the value of $\varsigma_{\rm stat,0}$. Also, note that when $N$ is very large i.e. $\lambda\gg1$, $\varsigma_{\rm net} \to \kappa\,\varsigma_{\rm sys}$.



\begin{table*}
\centering
\caption{Overview of results. `F' denotes the corresponding parameter has been ``fixed". `FwX' denotes the corresponding parameter has been ``fixed" with a Gaussian uncertainty of X\%. The number of injected events is 100000 for each dataset. For each dataset, $N$ is the number of detected binary BH events. $H_0$ is provided in km s$^{-1}$ Mpc$^{-1}$. $\Omega_{\rm m}$ is dimensionless. The recovered value corresponds to the peak of the probability distribution. The errors correspond to the 90\% credible interval.}
\renewcommand{\arraystretch}{1.6}
\begin{tabular}{cc|c|cc|cc|cc|c}
\hline
Case & Mock Dataset & No. of detected & \multicolumn{2}{c|}{Injected value} & \multicolumn{2}{c|}{Recovered value} & \multicolumn{2}{c|}{\% uncertainty} & \% uncertainty\\
\# & \# & BH events ($N$) & $H_0$ & $\Omega_{\rm m}$ & $H_0$ & $\Omega_{\rm m}$ & $H_0$ & $\Omega_{\rm m}$ & with $10^6$ events\\
\hline
I & 1 & 54850 & 67.3 & 0.3 & $67.8_{-2.2}^{+2.6}$ & F & 3.5 & - & \multirow{2}{*}{0.9\%}\\
II & 2 & 56455 & 73.5 & 0.3 & $73.3_{-2.2}^{+3.1}$ & F & 3.6 & - & \\
\hline
III & 3 & 55105 & 70.0 & 0.27 & F & $0.262_{-0.042}^{+0.032}$ & - & 14.1 & \multirow{2}{*}{3.6\%}\\
IV & 4 & 55767 & 70.0 & 0.32 & F & $0.310_{-0.042}^{+0.043}$ & - & 13.7 & \\
\hline
\multirow{3}{*}{V} & 5 & 55386 & 70.0 & 0.3 & $70.8_{-2.7}^{+2.2}$ & F & 3.5 & - & \multirow{3}{*}{0.9\%}\\
 & 6 & 55623 & 70.0 & 0.3 & $69.9_{-2.3}^{+3.0}$ & F & 3.8 & - & \\
 & 5+6 & 111009 & 70.0 & 0.3 & $70.7_{-1.7}^{+2.2}$ & F & 2.8 & - & \\
\hline
\multirow{3}{*}{VI} & 5 & 55386 & 70.0 & 0.3 & F & $0.291_{-0.038}^{+0.038}$ & - & 13.1 & \multirow{3}{*}{3.6\%}\\
 & 6 & 55623 & 70.0 & 0.3 & F & $0.289_{-0.047}^{+0.038}$ & - & 14.7 & \\
 & 5+6 & 111009 & 70.0 & 0.3 & F & $0.290_{-0.034}^{+0.028}$ & - & 10.7 & \\
\hline
\multirow{3}{*}{VII} & 5 & 55386 & 70.0 & 0.3 & $70.7_{-2.8}^{+2.6}$ & Fw3 & 3.8 & - & \multirow{3}{*}{1.6\%}\\
 & 6 & 55623 & 70.0 & 0.3 & $70.6_{-3.0}^{+2.5}$ & Fw3 & 3.9 & - & \\
 & 5+6 & 111009 & 70.0 & 0.3 & $70.7_{-2.4}^{+2.0}$ & Fw3 & 3.1 & - & \\
\hline
\multirow{3}{*}{VIII} & 5 & 55386 & 70.0 & 0.3 & Fw1 & $0.301_{-0.052}^{+0.031}$ & - & 13.8 & \multirow{3}{*}{7.0\%}\\
 & 6 & 55623 & 70.0 & 0.3 & Fw1 & $0.298_{-0.057}^{+0.033}$ & - & 15.1 & \\
 & 5+6 & 111009 & 70.0 & 0.3 & Fw1 & $0.295_{-0.045}^{+0.027}$ & - & 12.2 & \\
\hline
\multirow{3}{*}{IX} & 5 & 55386 & 70.0 & 0.3 & $70.8_{-3.3}^{+3.0}$ & Fw6 & 4.4 & - & \multirow{3}{*}{2.8\%}\\
 & 6 & 55623 & 70.0 & 0.3 & $70.0_{-3.0}^{+3.5}$ & Fw6 & 4.6 & - & \\
 & 5+6 & 111009 & 70.0 & 0.3 & $70.7_{-2.9}^{+2.5}$ & Fw6 & 3.8 & - & \\
\hline
\multirow{3}{*}{X} & 5 & 55386 & 70.0 & 0.3 & Fw2 & $0.302_{-0.064}^{+0.042}$ & - & 17.5 & \multirow{3}{*}{12.5\%}\\
 & 6 & 55623 & 70.0 & 0.3 & Fw2 & $0.301_{-0.075}^{+0.04}$ & - & 19.1 & \\
 & 5+6 & 111009 & 70.0 & 0.3 & Fw2 & $0.302_{-0.059}^{+0.033}$ & - & 15.2 & \\
\hline
\multirow{3}{*}{XI} & 5 & 55386 & 70.0 & 0.3 & $70.8_{-3.9}^{+3.6}$ & Fw9 & 5.3 & - & \multirow{3}{*}{4.1\%}\\
 & 6 & 55623 & 70.0 & 0.3 & $70.0_{-3.6}^{+4.1}$ & Fw9 & 5.5 & - & \\
 & 5+6 & 111009 & 70.0 & 0.3 & $70.4_{-3.4}^{+3.6}$ & Fw9 & 5.0 & - & \\
\hline
\multirow{3}{*}{XII} & 5 & 55386 & 70.0 & 0.3 & Fw3 & $0.302_{-0.086}^{+0.058}$ & - & 23.8 & \multirow{3}{*}{18.3\%}\\
 & 6 & 55623 & 70.0 & 0.3 & Fw3 & $0.301_{-0.087}^{+0.058}$ & - & 24.1 & \\
 & 5+6 & 111009 & 70.0 & 0.3 & Fw3 & $0.302_{-0.080}^{+0.046}$ & - & 20.9 & \\
 \hline
\multirow{3}{*}{XIII} & 5 & 55386 & 70.0 & 0.3 & $69.9_{-3.2}^{+4.6}$ & [0.25, 0.35] & 5.6 & - & \multirow{3}{*}{4.4\%}\\
 & 6 & 55623 & 70.0 & 0.3 & $69.7_{-3.3}^{+4.6}$ & [0.25, 0.35] & 5.7 & - & \\
 & 5+6 & 111009 & 70.0 & 0.3 & $69.6_{-2.7}^{+4.5}$ & [0.25, 0.35] & 5.2 & - & \\
\hline
\multirow{3}{*}{XIV} & 5 & 55386 & 70.0 & 0.3 & [65, 75] & $0.256_{-0.044}^{+0.123}$ & - & 32.6 & \multirow{3}{*}{25.1\%}\\
 & 6 & 55623 & 70.0 & 0.3 & [65, 75] & $0.250_{-0.046}^{+0.126}$ & - & 34.4 & \\
 & 5+6 & 111009 & 70.0 & 0.3 & [65, 75] & $0.250_{-0.038}^{+0.121}$ & - & 31.8 & \\
\hline
\end{tabular}
\label{overview}
\end{table*}

\section{Conclusions}

   \begin{enumerate}
      \item With $10^5$ binary BH events detected with ET as a standalone instrument, we should be able to constrain $H_0$ within 2.5\% uncertainty if other parameters of the SMOC are kept fixed.
      \item With the same number of binary BH events, we should be able to constrain $\Omega_{\rm m}$ within 10\% uncertainty if rest of the cosmological parameters are fixed.
      \item Applying the $1/\sqrt{N}$ rule for the decline in uncertainty, we estimate that 1\% uncertainty on $H_0$ may be achieved with $7\times10^5$ binary BH events, i.e. one year of ET's observation. This is consistent with earlier mock data challenge studies for ET \citep[e.g.][]{2021ApJ...908..215Y}. If $H_0$ is fixed to its true value, then $\Omega_{\rm m}$ may be estimated with 4\% uncertainty with the same number of events.
      \item The statistical uncertainty does not scale as $1/\sqrt{N}$ when a systematic uncertainty is present in the fixed parameter, and the uncertainty in the inferred parameter's value declines more slowly with the number of events.
      \item We provide a general formula to estimate the net uncertainty when there is a systematic uncertainty on the fixed parameter.
      \item The mock GW data was generated with a detector-centric coordinate system. If the rotation of earth were accounted for, the errors on the $d_L$ for all the events reduce significantly (relative errors are almost halved) \citep{2022PhRvD.106l3014S}, and hence the uncertainty would be considerably lower.
      \item The main bottleneck of this method is knowing the empirical intrinsic chirp mass distribution, $P(\mathcal{M})$. Moreover, any redshift variation of this mass spectrum may also affect the cosmological inference. The effect of $P(\mathcal{M})$ will be studied quantitatively in future.
      \item The intrinsic chirp mass distribution used in this work is derived from a specific model of isolated binary evolution. While there are also alternate models of binary evolution to consider which satisfy the local merger rate constraints, there is additionally the dynamical formation channel for compact binaries in dense star clusters. For a more comprehensive study, one should use a mass spectrum derived from both isolated and dynamical formation channels.
   \end{enumerate}

\begin{acknowledgements}
      The authors are supported by the OPUS grant 2023/49/B/ST9/02777 of the National Science Centre, Poland.
\end{acknowledgements}


%
%

\bibliographystyle{aa} 
\bibliography{aa} 

\end{document}